# Sizes of Ferroelectricity Appearance and Disappearence in Nanosized Hafnia-Zirconia: Landau-type Theory

Anna N. Morozovska[1*], Eugene A. Eliseev[2†], Sergei V. Kalinin[3‡], and Maksym V. Strikha[4,5§]

[1] Institute of Physics of the National Academy of Sciences of Ukraine, 46, Nauky Avenue, 03028 Kyiv, Ukraine

[2] Frantsevich Institute for Problems in Materials Science of the National Academy of Sciences of Ukraine, 3, str. Omeliana Pritsaka, 03142 Kyiv, Ukraine

[3] Department of Materials Science and Engineering, University of Tennessee, Knoxville, TN, 37996, USA

[4] Faculty of Radiophysics, Electronics and Computer Systems, Taras Shevchenko National University of Kyiv, Kyiv, Ukraine

[5] V. Lashkariov Institute of Semiconductor Physics, National Academy of Sciences of Ukraine, Kyiv, Ukraine

**Abstract**

Nanosized hafnia-zirconia $Hf_xZr_{1-x}O_2$ in the form of thin films, multilayers and heterostructures are indispensable silicon-compatible ferroelectric materials for advanced electronic memories and logic devices. The distinctive feature of nanoscale hafnia-zirconia are the critical sizes of ferroelectricity *appearance*, whereas the critical sizes of ferroelectricity *disappearance* exist in other ferroelectrics. Using the Landau-Ginzburg-Devonshire free energy functional with higher powers, trilinear and biquadratic couplings of polar, nonpolar and antipolar order parameters, we calculated analytically the strain-dependent critical sizes of the ferroelectricity appearance and disappearance, analyzed how the size effect and mismatch strains influence the phase diagrams and polarization switching barrier in epitaxial $HfO_2$ thin films and nano-islands with the out-of-plane spontaneous polarization. We have shown that the critical thickness/height of out-of-plane spontaneous polarization disappearance is determined by the size dependence of the depolarization field and correlation effects. The critical thickness/height of the ferroelectricity appearance is determined by the size dependence of the effective mismatch strain considering possible appearance of misfit dislocations and lateral relaxion of strains. Derived analytical expressions can be generalized for $Hf_xZr_{1-x}O_2$ solid solutions, providing that corresponding parameters of the free energy are known from the first principles calculations.

---

[*] Corresponding author: anna.n.morozovska@gmail.com

[†] E.A.E. and A.N.M. contributed equally

[‡] Corresponding author: sergei2@utk.edu

[§] Corresponding author: maksym.strikha@gmail.com

## I. INTRODUCTION

Nanosized hafnia-zirconia $Hf_xZr_{1-x}O_2$ (x ≥ 0.5), further abbreviated as HZO, in the form of thin films, multilayers and heterostructures are indispensable silicon-compatible ferroelectric materials for advanced electronic memories [1, 2] and logic devices [3, 4]. It was demonstrated that the transition from the bulk nonpolar monoclinic m-phase (space group *P2$_1$/c*) to the ferroelectric (FE) orthorhombic o-phase (space group *Pca2$_1$*) can happen in the nanoscale HZO with decrease in sizes below 30 nm [2-4]. For instance, HZO thin films reveal ferroelectric properties only when their thickness is less than the critical one, that does not exceed 30 nm, because the FE o-phase becomes metastable for larger thicknesses [5]. This experimental fact is opposite to the common trend for vast majority of ferroelectric films and particles, which have ferroelectric properties only when the film thickness or particle sizes are above the critical value, changing from several nanometers to sub-microns in dependence on the elastic strains and electric screening conditions at their surface. Thus, the distinctive feature of nanoscale hafnia-zirconia are the critical sizes of ferroelectricity *appearance*, whereas the critical sizes of ferroelectricity *disappearance* exist in other ferroelectrics. The critical size of ferroelectricity *disappearance* also exists in nanosized HZO at lower scales, because ferroelectricity is a collective phenomenon, which cannot exist in the volume smaller than the correlation volume.

At the same time, the physical mechanisms responsible for the emergence of ferroelectric and/or antiferroelectric properties in nanosized HZO are still unclear [6, 7]. According to the first principle theories, they are probably related with the dominant role of surface and size effects [5, 8], trilinear coupling between the "soft" polar and "hard" nonpolar modes enhanced by mismatch strains [9], hybridized [10] and/or nonpolar [11] phonon modes. Indeed, it was shown that the negative biquadratic coupling between polar and nonpolar modes [12] enhanced by large tensile strains the in-plane directions may induce the nonpolar mode instability in HfO$_2$ [13].

There are no comprehensive and widely accepted Landau-like models of either HfO$_2$ or ZrO$_2$, even the symmetry of their paraelectric and/or intermediate phases is still under debate. Despite the presence of the high temperature cubic (space group *Fm3m*) phase of HZO, the symmetrical considerations do not allow one to suppose that the polarization vector could cause the transition between the *Fm3m* and *Pca2$_1$* phases. Instead, the series of phase transitions should take place, and at least three order parameters (polar, antipolar and nonpolar) should be considered to achieve the correct description of the ferroelectric order in HfO$_2$, as it was shown by Jung and Birol [10] from the density functional theory (DFT) and Landau-type approach. Next, Jung and Birol [11] introduced the second-order dynamical charge along with phonon mode effective charge and applied these concepts to the HfO$_2$ and SrTiO$_3$ materials to show that the contribution of second-order dynamic charges to the HfO$_2$ polarization is rather large in contrast to perovskite ferroelectrics. They also obtained that the local polarization arising from the second-order effective charges is aligned opposite to the first-order polarization.

Using the DFT, Delodovici et al. [9] considered the transition from the high-symmetry nonpolar tetragonal phase *P4$_2$/nmc* (t-phase) to the polar o-phase (*Pca2$_1$*) in HfO$_2$. They noted that two Brillouin zone centered modes $\Gamma_{1+}$ (volume expansion) and $\Gamma_{4+}$ (shear strain) transform the *P4$_2$/nmc* phase to the nonpolar orthorhombic structure (space group *Ccce*). Considering the *Ccce* phase as the parent phase, they have shown that the combined action of three coupled modes (polar $\Gamma_{3-}$, nonpolar $Y_{2+}$ and antipolar $Y_{4-}$) are needed to achieve the energy minimum for the *Pca2$_1$* structure experimentally observed in HfO$_2$ films (see the black curve in **Fig. 1(a)**).

Also, Delodovici et al. [8, 9] studied the influence of biaxial elastic strains on ferroelectric properties of HfO$_2$ and its thin films. They obtained that compressive biaxial strains could induce ferroelectric polarization in rhombohedral r-phase of HfO$_2$, suggesting the absence of any critical thickness for the in-plane direction of the polarization [8]. At the same time, the stretching along the a-axis with simultaneous compression along other two axes (under the condition of volume conservation) is required to stabilize the out-of-plane ferroelectric polarization along the polar c-axis in HfO$_2$ films [9].

There exist experimental evidences about the decisive role of mismatch strains in appearance of the FE o-phase in epitaxial HZO thin films. For instance, Estandìa et al. [14] experimentally studied epitaxial HZO (x=0.5) films grown on different (001)-oriented cubic substrates and revealed that the appearance of the FE o-phase depends significantly on the mismatch strain. Later on, Estandìa et al. [15] studied (111)-oriented HZO films with (100)-LSMO electrodes and revealed the pronounced ferroelectric properties of the films, as well as report about an unusual epitaxy mechanism based on the observation of dislocations arrays of with short periodicities. Using the DFT and piezoelectric response force microscopy (PFM), Dutta et al. [16] explored the leading role of piezoelectricity in $HfO_2$ films, that allows us to relate the mismatch-induced effects with the high piezoelectric coupling in HZO. Zhou et al. [17] proposed a model based on the three order parameters, namely the tetragonal distortion mode, antipolar and polar modes, considering cubic fluorite structure as the parent phase. Tamura et al. [18] considered the influence of mismatch strains on the stability of FE o-phase in thin $HfO_2$ films.

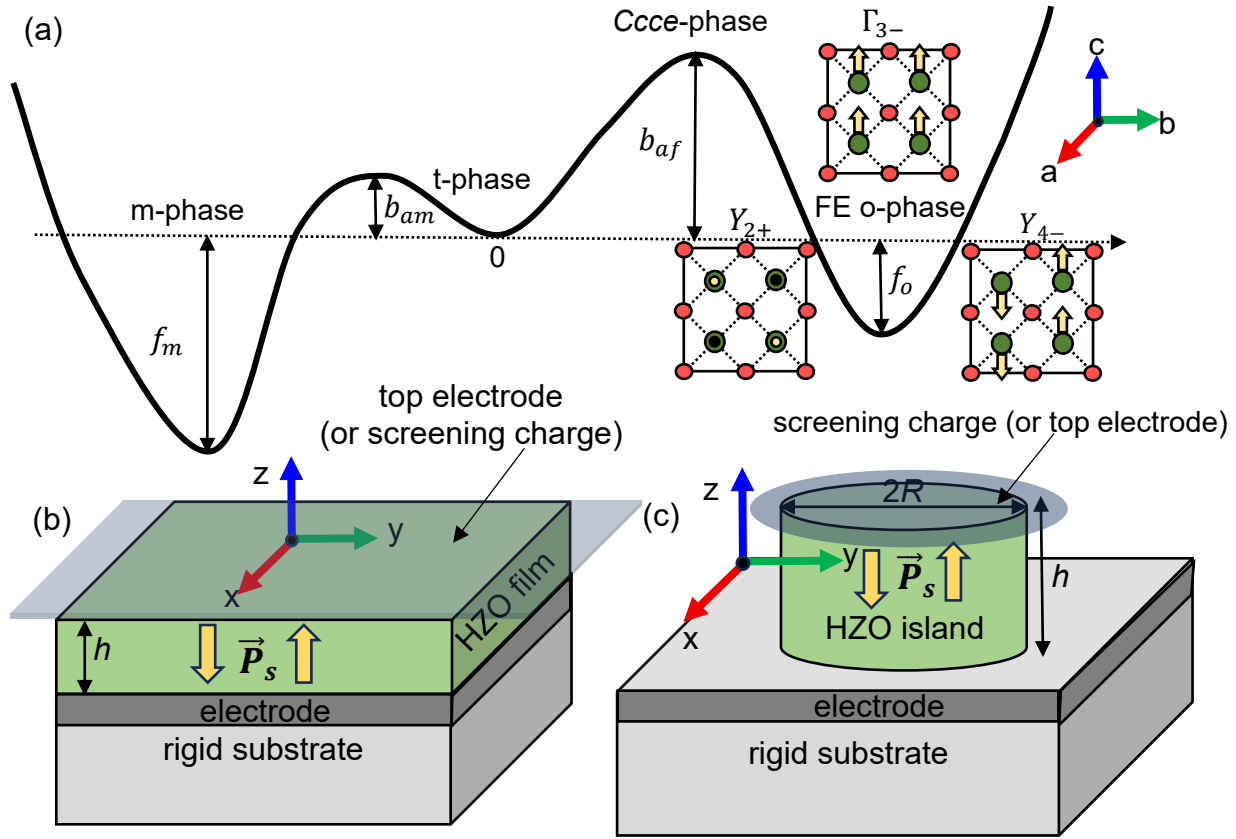

**FIGURE 1. (a)** Free energy density relief with the minima corresponding to the m-, t- and FE o-phases. Small insets show atomic displacements in the polar $\Gamma_{3-}$, nonpolar $Y_{2+}$ and antipolar $Y_{4-}$ modes, which exist in the FE o-phase of $HfO_2$. The barrier heights $b_{am}$ and $b_{af}$ determine activation energies of possible

t-m and t-o transitions. Geometry of epitaxial thin films **(b)** and nano-islands **(c)** of HZO subjected to epitaxial mismatch strain $u_m$. Two directions of the out-of-plane spontaneous polarization $\vec{P}_s$ in the FE o-phase of HZO are shown by the thick arrow. Z-axis coincides with the polar c-axis. The direction of a-axis is determined by the condensation of coupled $Y_{2+}$ and $Y_{4-}$ modes [9].

Important, that competing phases determine an indirect switching path of their spontaneous polarization $\vec{P}_s$ [19, 20]. According to *direct* experimental observations [21], the switching path of polarization in thin HZO films is indirect, because the transition from $+\vec{P}_s$ to $-\vec{P}_s$ goes through the nonpolar t-phase. Actually, Ooe et al. [21] observed that the pathways of 180- and 90-degree polarization rotation involve different nonpolar intermediate states with distinct spatial scales. For instance, the 180-degree domain walls between polar domains could be separated by either the t-phase *P4$_2$/nmc* or by the nonpolar o-phase *Pbcm*, depending on the wall orientation. The barrier height between different phases determines the activation energies of possible transitions, and so the most probable transition path (see e.g., **Fig. 1(a)**).

The situation with ferroelectricity existence and polarization switching is much more complicated in small HZO nanoparticles. Due to the small sizes in all three directions, it is hardly possible to separate the t-phase (space group *P4$_2$/nmc*) and o-phases (space groups *Pbca*, *Pbcm* and ferroelectric *Pca21*) using X-ray diffraction analysis, because corresponding peaks are very close, diffused and merge together [22]. Several experimental [23, 24] and theoretical [25, 26] works revealed a leading role of oxygen vacancies [27, 28] in the appearance and stabilization of the FE o-phase in nanoscale HZO. Recently, it was shown that the oxygen-deficient $Hf_xZr_{1-x}O_{2-y}$ nanoparticles could exhibit ferroelectric-like properties, such as a colossal dielectric response in a wide frequency range [29], as well as demonstrate resistive switching and pronounced charge accumulation [30].

Analysis of the spatial-temporal evolution of polarization in HZO thin films was performed in the framework of "effective" Landau-Ginzburg-Devonshire (LGD) thermodynamic approach [31, 32]. This approach incorporates the elements of the Kittel-type model [33] with polar and antipolar modes [34, 35] to the Landau-type free energy with effective parameters extracted either from the experiment [36] or from the DFT calculations. The influence of elastic strain and electrostriction coupling can be considered approximately by the renormalization of LGD coefficients [37]. Effective LGD approach [38], accomplished by Stephenson and Highland (SH) [39, 40] approach and corroborated by the DFT calculations [41], predicted that ferro-ionic states can be stable in spherical HZO nanoparticles of sizes 5 – 30 nm, being in agreement with recent experimental observations [29, 30].

In this work we propose the Landau-type theory for the determination of the ferroelectricity appearance and disappearance critical sizes in epitaxial HZO thin films and nano-islands with the out-of-plane spontaneous polarization. Our approach is based on results of Delodovici et al. [9], Jung and Birol [10], who determine the role of trilinear coupling, and Datta et al. results [16], which allowed us to extract the piezoelectric and electrostriction coupling coefficients. In result we constructed the LGD free energy functional considering higher powers, biquadratic and trilinear couplings of three order parameters (polar $\Gamma_{3-}$, nonpolar $Y_{2+}$ and antipolar $Y_{4-}$ modes), size and strain effects. Using the functional, we calculated analytically the strain-dependent critical sizes of the ferroelectricity appearance and disappearance, analyzed the mismatch strains influence and size effect of the phase diagrams and polarization switching barrier in epitaxial $HfO_2$ thin films and nano-islands, whose geometry is shown schematically in **Figs. 1(b)** and **1(c)**, respectively. Our approach for the determination of critical sizes can be generalized for HZO thin films and nanoparticles, provided that the relevant material parameters determining the LGD free energy coefficients can be found from first-principles calculations.

## II. THEORETICAL RESULTS

### A. Landau-Ginzburg-Devonshire Free Energy Functional of Nanosized Hafnia-Zirconia

Following Ref.[9], we consider the nonpolar t-phase as the reference "aristo-phase" of the HZO (x≥0.5). The free energy of a uniaxial HZO was expanded with respect to the powers of the order parameters, introduced as the dimensionless amplitudes $Q_{\Gamma 3}$, $Q_{Y2}$ and $Q_{Y4}$ of the corresponding phonon modes normalized to their equilibrium values.

The full form of the expression for the LGD free energy bulk density $f_{bulk}$, which is the expansion up to the 8-th powers of $Q_{\Gamma 3}$, $Q_{Y2}$ and $Q_{Y4}$ considering the t-symmetry of aristo-phase is derived in **Appendix S1** [42]. The compact form of $f_{bulk}$ is the sum of the 2-4-6-8 powers and 1-5-7 powers of the three phonon modes ($f_{bl}$ and $f_{tr}$), elastic and striction energy contributions ($f_{est}$), and the gradient energy of the order parameters ($f_{grad}$):

$$f_{bulk} = f_{bl} + f_{tr} + f_{est} + f_{grad}, \qquad (1a)$$

$$f_{bl} = \beta_i Q_i^2 + \delta_{ij} Q_i^2 Q_j^2 + \eta_{ijk} Q_i^2 Q_j^2 Q_k^2 + \xi_{ijkl} Q_i^2 Q_j^2 Q_k^2 Q_l^2, \qquad (1b)$$

$$f_{tr} = (\gamma + \epsilon_i Q_i^2 + \zeta_{ij} Q_i^2 Q_j^2) Q_{\Gamma 3} Q_{Y2} Q_{Y4}, \tag{1c}$$

$$f_{est} = \frac{1}{2} c_{ijkl} u_{ij} u_{kl} - \tilde{q}_{ijkl} u_{ij} Q_k Q_l - (1 + v_i Q_i^2 + k_{ij} Q_i^2 Q_j^2) \tilde{r}_{ijklm} u_{ij} Q_k Q_l Q_m - \tilde{z}_{ijklmn} u_{ij} Q_k Q_l Q_m Q_n + \widetilde{w}_{ijklmn} u_{ij} u_{kl} Q_m Q_n, \tag{1d}$$

$$f_{grad} = \frac{1}{2} g_{ijkl} \frac{\partial Q_k}{\partial x_i} \frac{\partial Q_l}{\partial x_j}. \tag{1e}$$

Here the subscripts $i, j, k, l, m, n$ ... either designate the modes $\Gamma_{3-}$, $Y_{2+}$ and $Y_{4-}$, or are the Cartesian indexes coupled to the strains $u_{ij}$, or to the Cartesian coordinates $x_i$. For instance, $Q_i$ stands for $Q_{\Gamma 3}$, $Q_{Y2}$ or $Q_{Y4}$. The summation rule is performed over repeated subscripts. Following Ref.[9], we pay special attention to the strong trilinear coupling of the $Q_{\Gamma 3}$, $Q_{Y2}$ and $Q_{Y4}$ modes, which energy $f_{tr}$ is proportional to the product $Q_{\Gamma 3} Q_{Y2} Q_{Y4}$, since the coupling can stabilize the FE o-phase.

Nonzero components of $\beta_i$, $\gamma$, $\delta_{ij}$, $\epsilon_i$, $\eta_{ijk}$, $\zeta_{ij}$ and $\xi_{ijkl}$ used in calculations for HfO$_2$ are listed in **Tables S1-S2** in **Appendix S1** [42]. $c_{ijkl}$ are elastic stiffnesses, $u_{ij}$ are elastic strains, $\tilde{q}_{ijkl}$, $\tilde{z}_{ijklmn}$ and $\widetilde{w}_{ijklmn}$ are the components of the second-order and higher-order striction *stress* tensors; $\tilde{r}_{ijklm}$ are the tensor of trilinear striction; $g_{ijkl}$ are the components of the gradient energy tensor.

The ferroelectric polarization components $P_i$, both spontaneous and induced by external electric field $E_i$, contribute to the electric energy $f_{el}$:

$$f_{el} = -P_i E_i - \frac{1}{2} P_i E_i^d. \tag{2a}$$

Here the subscript $i = 1, 2, 3$ and $E_i^d$ is the depolarization field, which should be determined from electrostatic equations in a self-consistent way. Following Refs. [9, 10, 11], the spontaneous polarization $\vec{P}_S$ is proportional to the amplitude $Q_{\Gamma 3}$ of the $\Gamma_{3-}$ mode. Let us assume that:

$$P_3 = \frac{Z_B^* d}{V_{f.u.}} Q_{\Gamma 3} \approx P_S Q_{\Gamma 3}, \tag{2b}$$

where $Z_B^*$ is the effective Bader charge [11], $V_{f.u.}$ is the formula unit (f.u.) volume and $d$ is the elementary displacement corresponding to the polar $\Gamma_{3-}$ mode. The spontaneous polarization value $P_S \approx 54.8$ µC/cm$^2$ was calculated in Ref. [9] for the hypothetic bulk HfO$_2$. At the same time the amplitude of the maximal atomic displacement $d$ was reported as 0.284, 0.278 and 0.268 Å for the polar, nonpolar and antipolar modes, respectively.

Substitution of $P_3$ in Eqs.(1) instead of $Q_{\Gamma 3}$, leads to the rescaling of the coefficients $\beta_i$, $\gamma$, $\delta_{ij}$, $\epsilon_i$, $\eta_{ijk}$, $\zeta_{ij}$, $\xi_{ijkl}$, $q_{ijkl}$ and $g_{ijkl}$ proportional to the powers of the factor $\frac{V_{f.u.}}{Z_B^* d} \approx \frac{1}{P_S}$. In particular, the biquadratic 2-4-6-8 coupling coefficients $\tilde{\beta}_{\Gamma 3}$, $\tilde{\delta}_{\Gamma 3}$, $\tilde{\eta}_{\Gamma 3}$, $\tilde{\xi}_{\Gamma 3}$ and the trilinear coupling coefficient $\tilde{\gamma}$, responsible for the behavior of the polar mode $\Gamma_{3-}$, have the form:

$$\tilde{\beta}_{\Gamma 3} = \frac{\beta_{\Gamma 3}}{P_S^2}, \quad \tilde{\delta}_{\Gamma 3} = \frac{\delta_{\Gamma 3}}{P_S^4}, \quad \tilde{\eta}_{\Gamma 3} = \frac{\eta_{\Gamma 3}}{P_S^6}, \quad \tilde{\xi}_{\Gamma 3} = \frac{\xi_{\Gamma 3}}{P_S^8}, \quad \tilde{\gamma} = \frac{\gamma}{P_S}. \tag{3}$$

In what follows we consider nanosized HZO, such as thin films epitaxially grown at a rigid substrate and epitaxial nano-islands, whose geometry are shown schematically in **Fig. 1(b)** and **1(c)**. Following Refs. [5, 9, 10, 18], we consider the nanosized HZO as a uniaxial ferroelectric material with the out-of-plane spontaneous polarization, whose polar z-axis is normal to the film surfaces. Hereinafter we regard that surfaces of nanosized HZO are covered with ideally conducting or semiconducting electrodes, or with ionic-electronic screening charges. The screening charges may originate from the oxygen vacancies accumulation by the surface.

Next, we assume that the surface/interface energy excess $F_s$ of the HZO is the biquadratic form of $Q_i$:

$$F_s = \int \alpha_i Q_i^2 dS = F_{SP} + F_{SY}, \tag{4a}$$

$$F_{SP} = \int (\alpha_{\Gamma 3} P_3^2 + \alpha_{\Gamma 3Y2} P_3 Q_{Y2} + \alpha_{\Gamma 3Y4} P_3 Q_{Y4}) dS, \tag{4b}$$

$$F_{SY} = \int (\alpha_{Y2} Q_{y2}^2 + \alpha_{Y4} Q_{Y4}^2 + \alpha_{Y2Y4} Q_{Y2} Q_{Y4}) dS. \tag{4c}$$

Since the m-phase is stable in the bulk, the energy $F_{SY} \geq 0$ at arbitrary values of $Q_{Y2}$ and $Q_{Y4}$, that is true for $\alpha_{Y2} > 0$, $\alpha_{Y4} > 0$ and $2\sqrt{\alpha_{Y2}\alpha_{Y4}} > -\alpha_{Y2Y4}$. Since the FE o-phase can be stable in the nanosized HZO only, the surface energy $F_{SP}$, related to the appearance of the polar order, may be negative. Following the surface stability conditions, the total surface energy $F_s$ should be positive. In what follows we consider the simplest diagonalized form of the surface energy, with $\alpha_{\Gamma 3Y2} = \alpha_{\Gamma 3Y4} = \alpha_{Y2Y4} = 0$ and $\alpha_{Y2} > 0$, $\alpha_{Y4} > 0$. The case $\alpha_{\Gamma 3} < 0$ supports the stability of the polar $\Gamma_{3-}$ mode at the nanoscale and contributes to the surface-induced phase transition. However, since the microscopic mechanisms, which may be responsible for $\alpha_{\Gamma 3} < 0$, are not considered, we put $\alpha_{\Gamma 3} \geq 0$ in numerical calculations.

The free energy of the nanosized HZO corresponding to the FE o-phase, $F_{o-phase}$, is the volume integral of the LGD free energy bulk $f_{bulk}$ given by Eq.(1), electric energy $f_{el}$ given by Eq.(2) and the surface/interface energy $F_s$ given by Eqs.(4):

$$F_{o-phase} = \int (f_{bl} + f_{tr} + f_{est} + f_{grad} + f_{el})dV + F_s. \tag{5a}$$

The FE o-phase can be stable in the nanosized HZO when its free energy $F_{o-phase}$ is smaller than the energy of the m-phase, $F_m = \int f_m dV$. The critical sizes of the ferroelectricity appearance in nanosized HZO can be estimated from the condition

$$F_{o-phase} = F_m. \tag{5b}$$

The density of a bulk HfO$_2$ m-phase energy $f_m$, counted from the t-phase, is about -92 meV/f.u. [9].

Due to elastic strains, surface tension, polarization gradient energy and depolarization field, coefficients in the free energy (5) become "renormalized". Their renormalization depends on the geometry, mismatch strains, size effects and screening conditions (see e.g., Refs. [31, 32, 38]). In next sections we consider the cases of epitaxial HZO thin films and nano-islands with a single-domain out-of-plane spontaneous polarization.

### B. Epitaxial Thin Films of Hafnia-Zirconia

Next let us consider a single-domain HZO thin film epitaxially grown on a rigid substrate (corresponding geometry is shown in **Fig. 1(b)**). The epitaxial biaxial strain $u_m$ induced by the film-substrate lattice constants mismatch is $u_m = \frac{b-a}{a}$, where $a$ and $b$ are the film and substrate in-plane lattice constants, respectively. The depolarization field inside the single-domain film with an out-of-plane polarization can be estimated as [43]:

$$E_3(z) = \frac{\bar{P}_3 - P_3(z)}{\varepsilon_0 \varepsilon_b} - \frac{\bar{P}_3}{\varepsilon_0 \varepsilon_b} \frac{\lambda_{eff}}{\lambda_{eff} + h/\varepsilon_b}, \tag{6}$$

where $P_3(z)$ is the out-of-plane polarization directed along z-axis, $\bar{P}_3$ is the z-averaged polarization; $h$ is the film thickness, $\lambda_{eff} = \lambda_1/\varepsilon_{g1} + \lambda_2/\varepsilon_{g2}$ is the effective screening length corresponding to conducting or semiconducting top (subscript "1") and bottom (subscript "2") electrode(s) with the dielectric constants $\varepsilon_{gi}$; $\varepsilon_b$ is a background dielectric permittivity [44] of the film, $\varepsilon_0$ is a universal dielectric constant. Domain formation can emerge in thin films under imperfect screening conditions [45], but in this work we would like to focus on the strain-induced effects emerging in single-domain epitaxial films and nano-islands. This is possible for very small effective screening lengths, $\lambda_{eff} \leq 1$ Å [46], when the depolarization field becomes negligibly small and the domain formation is not energetically favorable. Despite $\lambda_{eff}$ is much smaller than the lattice constant in the case, its value may be quite realistic, because $\lambda_{eff}$ is related with the real charge-surface separation $\lambda_i$ as $\lambda_{eff} \cong \lambda_i/\varepsilon_{gi}$. Since the dielectric constants $\varepsilon_{gi}$ can be large enough, the separation $\lambda_i$ can appear higher than the lattice constant. The limiting case $\lambda_{eff} \to 0$ corresponds to the ideally conducting electrodes. The case of large $\lambda_{eff} > 1$ nm may correspond to the sluggish ionic-electronic screening charges (instead of the top electrode).

Using the approach evolved in Refs. [47, 48] for single-domain epitaxial ferroelectric films, the free energy of the FE o-phase of a single-domain epitaxial HZO thin film is given by approximate expression:

$$F_{o-phase} \approx \tilde{\beta}_f \bar{P}_3^2 + \tilde{\beta}_{Y2}\bar{Q}_{Y2}^2 + \tilde{\beta}_{Y4}\bar{Q}_{Y4}^2 + \tilde{\delta}_f \bar{P}_3^4 + \tilde{\delta}_{Y2}\bar{Q}_{Y2}^4 + \tilde{\delta}_{Y4}\bar{Q}_{Y4}^4 + \tilde{\delta}_{Y24}\bar{Q}_{Y2}^2\bar{Q}_{Y4}^2 +$$
$$\tilde{\eta}_f \bar{P}_3^6 + \tilde{\xi}_f \bar{P}_3^8 + \tilde{\gamma}_f \overline{Q_{Y2}Q_{Y4}P_3} - \overline{P_3 E_3}, \quad (7a)$$

where renormalized coefficients $\tilde{\beta}_{\Gamma 3}, \tilde{\delta}_{\Gamma 3}, \tilde{\eta}_{\Gamma 3}, \tilde{\xi}_{\Gamma 3}$ and $\tilde{\gamma}_f$ have the form:

$$\tilde{\beta}_f = \tilde{\beta}_{\Gamma 3} + \frac{\lambda_{eff}}{2\varepsilon_0 \varepsilon_b (h+\lambda_{eff})} + \frac{g}{h\Lambda_P + h^2/\pi} - u_m^* \left( \tilde{q}_{13} + \tilde{q}_{23} - \frac{c_{13}+c_{23}}{c_{33}} \tilde{q}_{33} \right), \quad (7b)$$

$$\tilde{\delta}_f = \tilde{\delta}_{\Gamma 3} - \frac{\tilde{q}_{33}^2}{2c_{33}} - u_m^* \left( \tilde{z}_{133} + \tilde{z}_{233} - \frac{c_{13}+c_{23}}{c_{33}} \tilde{z}_{333} \right), \quad (7c)$$

$$\tilde{\eta}_f = \tilde{\eta}_{\Gamma 3} - \frac{\tilde{q}_{33}\tilde{z}_{333}}{c_{33}}, \quad (7d)$$

$$\tilde{\xi}_f = \tilde{\xi}_{\Gamma 3} - \frac{\tilde{z}_{333}^2}{2c_{33}}, \quad (7e)$$

$$\tilde{\gamma}_f = \tilde{\gamma} - u_m^* \left[ \tilde{r}_{133} + \tilde{r}_{233} - \frac{c_{13}+c_{23}}{c_{33}} \tilde{r}_{333} \right]. \quad (7f)$$

Here $\bar{P}_3$, $\bar{Q}_{Y2}$ and $\bar{Q}_{Y4}$ are the averaged order parameters; $c_{ijkl}$ are elastic compliances, $\tilde{q}_{ijkl}, \tilde{r}_{ijklm}$ and $\tilde{z}_{ijkimn}$ are the second order and the higher-order electrostriction *stress* tensor components, written in Voight or mixed notations. A possible presence of misfit dislocations leads to the relaxation of the "seeding" mismatch strain $u_m$, and it should be substituted with the "effective" misfit strain

$$u_m^* = u_m \left( 1 - \exp\left[ -\frac{h_d}{h} \right] \right), \quad (7g)$$

where $u_1 = u_2 = u_m$ are the components of the seeding biaxial mismatch strain, $h_d$ is the critical thickness of dislocation appearance [49].

The second term in Eq.(7b) originates from the depolarization field. The third term originates from the polarization gradient energy [50]; $g = \frac{1}{2}(g_{3311} + g_{3322})$ is the combination of the gradient energy tensor components, and $\Lambda_P = \frac{\alpha_{\Gamma 3}}{g}$ is the polarization extrapolation length at the film surfaces [51]. Since $\lambda_{eff} \geq 0$ and $\Lambda_P \geq 0$, these two terms are always positive and thus can only increase the value of $\tilde{\beta}_f$, thereby worsening the ferroelectric properties of the HZO film. The last term in Eq.(5b) is the strain-induced renormalization of $\tilde{\beta}_f$ proportional to $u_m$. It has different trends for tensile ($u_m > 0$) and compressive ($u_m < 0$) mismatch strains and thus can enhance or worsen the ferroelectric properties of the HZO film.

The strain-induced renormalization of the coefficients $\tilde{\delta}_f$ and $\tilde{\gamma}_f$ is linearly $u_m$-dependent according to Eqs.(7b) and (7f), and so $\tilde{\delta}_f$ and $\tilde{\gamma}_f$ can change their sign due to the mismatch strain. At the same time, clamping to substrate changes the nonlinear coefficients $\tilde{\eta}_r$ and $\tilde{\xi}_r$ in a "global" mismatch-independent way (see Eqs.(7d) and (7e)). Thus, a film clamping to a rigid substrate can make the potential well corresponding to the FE o-phase more shallow or deeper and shift its position due to the change of the spontaneous polarization magnitude. The change of the potential well depth is defined by the magnitudes and signs of the renormalized coefficients in Eqs.(7a). Note that $\tilde{\beta}_f = \tilde{\beta}_{\Gamma 3} + \frac{\lambda_{eff}}{2\varepsilon_0 \varepsilon_b (h+\lambda_{eff})} + \frac{2g}{h\Lambda_P + h^2/4}$, $\tilde{\delta}_f = \tilde{\delta}_{\Gamma 3}$, $\tilde{\eta}_f = \tilde{\eta}_{\Gamma 3}$, $\tilde{\xi}_f = \tilde{\xi}_{\Gamma 3}$ and $\tilde{\gamma}_f = \tilde{\gamma}$ for a free-standing single-domain HZO film.

Next, the critical thickness $h_{cr}$ of the ferroelectricity appearance in strained thin HZO films can be estimated from the condition $F_{o-phase} = F_m$. The energy $F_{o-phase}$ is given by approximate expression (5a), where the order parameters $\bar{P}_3$, $\bar{Q}_{Y2}$ and $\bar{Q}_{Y4}$, averaged over the film thickness, satisfy the coupled equations

$$2\tilde{\beta}_f \bar{P}_3 + 4\tilde{\delta}_f \bar{P}_3^3 + 6\tilde{\eta}_f \bar{P}_3^5 + 8\tilde{\xi}_f \bar{P}_3^7 = \bar{E}_3 - \tilde{\gamma}_f \bar{Q}_{Y2} \bar{Q}_{Y4}, \tag{8a}$$

$$2\tilde{\beta}_{Y2} \bar{Q}_{Y2} + 4\tilde{\delta}_{Y2} \bar{Q}_{Y2}^3 + 2\tilde{\delta}_{Y24} \bar{Q}_{Y4}^2 \bar{Q}_{Y2} = -\tilde{\gamma}_f \bar{P}_3 \bar{Q}_{Y4}, \tag{8b}$$

$$2\tilde{\beta}_{Y4} \bar{Q}_{Y4} + 4\tilde{\delta}_{Y4} \bar{Q}_{Y4}^3 + 2\tilde{\delta}_{Y24} \bar{Q}_{Y2}^2 \bar{Q}_{Y4} = -\tilde{\gamma}_f \bar{P}_3 \bar{Q}_{Y2}. \tag{8c}$$

For small strength of the trilinear coupling the amplitudes of the antipolar and nonpolar modes are $\bar{Q}_{Y2s} \approx \pm \sqrt{-\frac{2\tilde{\beta}_{Y2}\tilde{\delta}_{Y4} - \tilde{\beta}_{Y4}\tilde{\delta}_{Y24}}{4\tilde{\delta}_{Y2}\tilde{\delta}_{Y4} - \tilde{\delta}_{Y24}^2}}$ and $\bar{Q}_{Y4s} \approx \pm \sqrt{-\frac{2\tilde{\beta}_{Y4}\tilde{\delta}_{Y2} - \tilde{\beta}_{Y2}\tilde{\delta}_{Y24}}{4\tilde{\delta}_{Y2}\tilde{\delta}_{Y4} - \tilde{\delta}_{Y24}^2}}$. In this case the out-of-plane polarization obeys the equation

$$2\tilde{\beta}_f \bar{P}_3 + 4\tilde{\delta}_f \bar{P}_3^3 + 6\tilde{\eta}_f \bar{P}_3^5 + 8\tilde{\xi}_f \bar{P}_3^7 = \bar{E}_3 \pm \tilde{\gamma}_f \frac{\sqrt{(2\tilde{\beta}_{Y2}\tilde{\delta}_{Y4} - \tilde{\beta}_{Y4}\tilde{\delta}_{Y24})(2\tilde{\beta}_{Y4}\tilde{\delta}_{Y2} - \tilde{\beta}_{Y2}\tilde{\delta}_{Y24})}}{4\tilde{\delta}_{Y2}\tilde{\delta}_{Y4} - \tilde{\delta}_{Y24}^2}. \tag{9}$$

It is seen that the spontaneous polarization can be incipient, namely $\bar{P}_s \approx \mp \frac{\tilde{\gamma}_f}{2\tilde{\beta}_f} \frac{\sqrt{(2\tilde{\beta}_{Y2}\tilde{\delta}_{Y4} - \tilde{\beta}_{Y4}\tilde{\delta}_{Y24})(2\tilde{\beta}_{Y4}\tilde{\delta}_{Y2} - \tilde{\beta}_{Y2}\tilde{\delta}_{Y24})}}{4\tilde{\delta}_{Y2}\tilde{\delta}_{Y4} - \tilde{\delta}_{Y24}^2}$ for positive $\tilde{\beta}_f$, $\tilde{\delta}_f$, $\tilde{\eta}_f$ and $\tilde{\xi}_f$. For negative $\tilde{\beta}_f$, a reversable ferroelectric polarization could appear. The values of $\bar{P}_s$ can be estimated from Eq.(9) at $\bar{E}_3 = 0$. In the simplest case, when $\tilde{\delta}_f > 0$, $\tilde{\eta}_f$ and $\tilde{\xi}_f$ are negligibly small, the condition $\tilde{\beta}_f = 0$ determines the borderline between incipient and reversable ferroelectric states. However, the condition $\tilde{\beta}_f \leq 0$ is a necessary but not sufficient condition of the polar phase stability. The sufficient condition is

$$\tilde{\beta}_f \bar{P}_s^2 + \tilde{\delta}_f \bar{P}_s^4 + \tilde{\eta}_f \bar{P}_s^6 + \tilde{\xi}_f \bar{P}_s^8 + \tilde{\gamma}_f \bar{Q}_{Y2s} \bar{Q}_{Y4s} \bar{P}_s \leq F_m - F_{QY}, \tag{10a}$$

where $F_{QY}$ is the average energy density of antipolar and nonpolar modes energy:

$$F_{QY} = \tilde{\beta}_{Y2} \bar{Q}_{Y2s}^2 + \tilde{\beta}_{Y4} \bar{Q}_{Y4s}^2 + \tilde{\delta}_{Y2} \bar{Q}_{Y2s}^4 + \tilde{\delta}_{Y4} \bar{Q}_{Y4s}^4 + \tilde{\delta}_{Y24} \bar{Q}_{Y2s}^2 \bar{Q}_{Y4s}^2. \tag{10b}$$

To derive relatively simple equation for the critical thickness from Eqs.(10), we should assume that the spontaneous polarization $\bar{P}_s$ relatively weakly depends on the mismatch strain and film thickness in the "deep" FE o-phase. Using these assumptions, an approximate transcendental equation for the critical thickness $h_{cr}$ acquires the form:

$$\tilde{\beta}_{\Gamma 3} + \frac{\lambda_{eff}}{2\varepsilon_0 \varepsilon_b (h_{cr} + \lambda_{eff})} + \frac{g}{h_{cr} \Lambda_P + h_{cr}^2/\pi} - u_m \left(1 - \exp\left[-\frac{h_d}{h_{cr}}\right]\right) \left(\tilde{q}_{13} + \tilde{q}_{23} - \frac{c_{13} + c_{23}}{c_{33}} \tilde{q}_{33}\right) = -\Delta_s,$$

$$\tag{11}$$

where we introduced the positive parameter $\Delta_s$ as:

$$\Delta_s = \tilde{\delta}_f \bar{P}_s^2 + \tilde{\eta}_f \bar{P}_s^4 + \tilde{\xi}_f \bar{P}_s^6 + \tilde{\gamma}_f \frac{\bar{Q}_{Y2s} \bar{Q}_{Y4s}}{\bar{P}_s} + \frac{F_{QY} - F_m}{\bar{P}_s^2}. \tag{12}$$

Assuming that the contribution of higher-order electrostriction components $\tilde{z}_{ijk}$ and $\tilde{r}_{ijk}$ are small, we can neglect the mismatch-dependent terms proportional in the coefficients $\tilde{\delta}_f$ and $\tilde{\gamma}_f$, given by Eqs.(7c) and (7f), respectively. Considering this and other abovementioned assumptions, the parameter $\Delta_s$ can be regarded as mismatch independent.

Note that all figures, discussed below, are calculated using exact Eqs.(7)-(8), because the accuracy of approximate Eqs.(9)-(11) is not high enough near the boundaries between different phases. However, Eqs.(9)-(11) well describe qualitatively the polarization behavior and critical sizes in thin HZO films.

Phase state of the epitaxial HfO$_2$ film as a function of mismatch strain $u_m$ and thickness $h$ is shown in **Fig. 2** for several typical values of the effective screening length: $\lambda_{eff} = 0$ (that corresponds to the ideally conducting electrodes), $\lambda_{eff} = 0.2$ Å (that corresponds to the semiconducting electrodes with high conductivity [46]) and $\lambda_{eff} = 1$ Å (that corresponds to the semiconducting electrodes or to the ionic-electronic screening charge layer [38] at the film surface).

It is seen that the FE o-phase can be stable at compressive mismatch strains $u_m < u_{min}$ only, where the minimal strain $u_{min}$ changes from $-0.25$ % at $\lambda_{eff} = 0$ to $-2$ % at $\lambda_{eff} = 1$ Å due to the size effect of depolarization field (see Eq.(6)). The conclusion about an urgency to apply compressive mismatch strains to induce the FE o-phase in HfO$_2$ films agrees with DFT results [9, 18]. The area of the FE o-phase region stability decreases strongly with increase in $\lambda_{eff}$, being replaced by the nonpolar m-phase (compare the first, the second and the third columns in **Fig. 2**). The m-phase becomes reentrant for $\lambda_{eff} > 0$. The FE o-phase disappears completely for $\lambda_{eff} > \lambda_{max}$, where $\lambda_{max}$ increases with increase in the maximal compressive mismatch strain $u_{max}$. In particular, $\lambda_{max} \approx 2$ Å at $u_{max} \approx -3$ %. Since the domain formation should be considered in the case $\lambda_{max} > 1$ Å, which may decrease the critical thickness of ferroelectricity disappearance, the case of larger $\lambda_{max}$ and $u_{max}$ will be considered elsewhere.

Color scale in **Figs. 2(a)-2(c)** is the absolute value of spontaneous polarization $\bar{P}_s$ in the deepest potential well of the free energy (7). The sharp boundary between the FE o-phase (with $\bar{P}_s > 0$) and nonpolar m-phase (with $\bar{P}_s = 0$) is the first order phase transition curve describing the dependence of the critical thickness $h_{cr}$ on the mismatch strain $u_m$. As follows from Eq.(11), the strain-dependent critical thickness of the ferroelectricity appearance ($h_{cr}^{max}$) is determined by the size dependence of the effective mismatch strain $u_m^*$ (proportional to the function $u_m \left(1 - \exp\left[-\frac{h_d}{h}\right]\right)$ according to Eq.(7g)). Due to dislocation appearance, the power decay of mismatch strains occurs in films of thickness $h > h_d$, namely $u_m^* \sim u_m \frac{h_d}{h}$. Thus, the estimation $h_{cr}^{max} \leq 10 h_d$ may be valid. The critical thickness of the ferroelectricity disappearance ($h_{cr}^{min}$) is determined by the size dependence of the depolarization field (proportional to the function $\frac{\lambda_{eff}}{2\varepsilon_0 \varepsilon_b (h + \lambda_{eff})}$ according to Eq.(6)) and correlation effects (proportional to the function $\frac{g}{h \Lambda_P + h^2/\pi}$). Hence, the estimations $h_{cr}^{min} \gg \lambda_{eff}$ and $h_{cr}^{min} \gg \sqrt{g \varepsilon_0 \varepsilon_b}$ should be valid. Dependence of the film critical thicknesses, $h_{cr}^{max}$ and $h_{cr}^{min}$, on the mismatch strain $u_m$ is given by the boundary between the FE o-phase and the m-phase (see white circles in **Figs. 2(a)-2(c)**). For ideal screening conditions, when $\lambda_{eff} = 0$, the FE o-phase is thermodynamically stable in HZO films with thickness $h < h_{cr}^{max}$ (see **Fig. 2(a)**). In this case the minimal thickness is limited by the correlation thickness being equal to several lattice constants. For incomplete screening conditions, when $0 < \lambda_{eff} < \lambda_{max}$, the FE o-phase is thermodynamically stable in HZO films with thickness $h_{cr}^{min} <$ $h < h_{cr}^{max}$ (see white circles in **Figs. 2(b)** and **2(c)**); at that $\lambda_{max}$ is determined by the maximal mismatch strain $u_{max}$. The thicknesses $h_{cr}^{min}$ and $h_{cr}^{max}$ become equal at $\lambda_{eff} \to \lambda_{max}$ and disappear at $\lambda_{eff} > \lambda_{max}$.

Color scale in **Figs. 2(d)-2(f)** is the absolute value of the nonpolar mode $\bar{Q}_{Y2s}$ in the deepest potential well of the free energy. The difference of the antipolar and nonpolar modes magnitudes, $|\bar{Q}_{Y4s}| - |\bar{Q}_{Y2s}|$, is shown in **Figs. 2(g)-2(i)**. It is seen that the difference $|\bar{Q}_{Y4s}| - |\bar{Q}_{Y2s}|$ is very small: it does not exceed 0.5 pm, that is 50 times smaller than the maximal value of $\bar{Q}_{Y2s}$ ($\approx 25$ Å). The small value of the difference means that the approximation $|\bar{Q}_{Y4s}| \approx |\bar{Q}_{Y2s}|$ is valid with high accuracy in the minimum of the free energy (7). From a physical standpoint, the approximate equality $|\bar{Q}_{Y4s}| \approx |\bar{Q}_{Y2s}|$ means a negligible anisotropy of the free energy potential well as a function of $|\bar{Q}_{Y4s}|$ and $|\bar{Q}_{Y2s}|$.

The deepest minimum of the free energy density, $f_{o-phase}^{min}$, as a function of $u_m$ and $h$, calculated for $\lambda_{eff} = 0$, 0.2 Å and 1 Å, are shown in **Figs. 3(g), 3(k)** and **3(l)**, respectively. A red background in the figure corresponds to the bulk m-phase with the energy density $f_m = -92$ meV/f.u. [9]. The first order phase transition between the o-phase and m-phase corresponds to the condition $f_{o-phase}^{min} = f_m$.

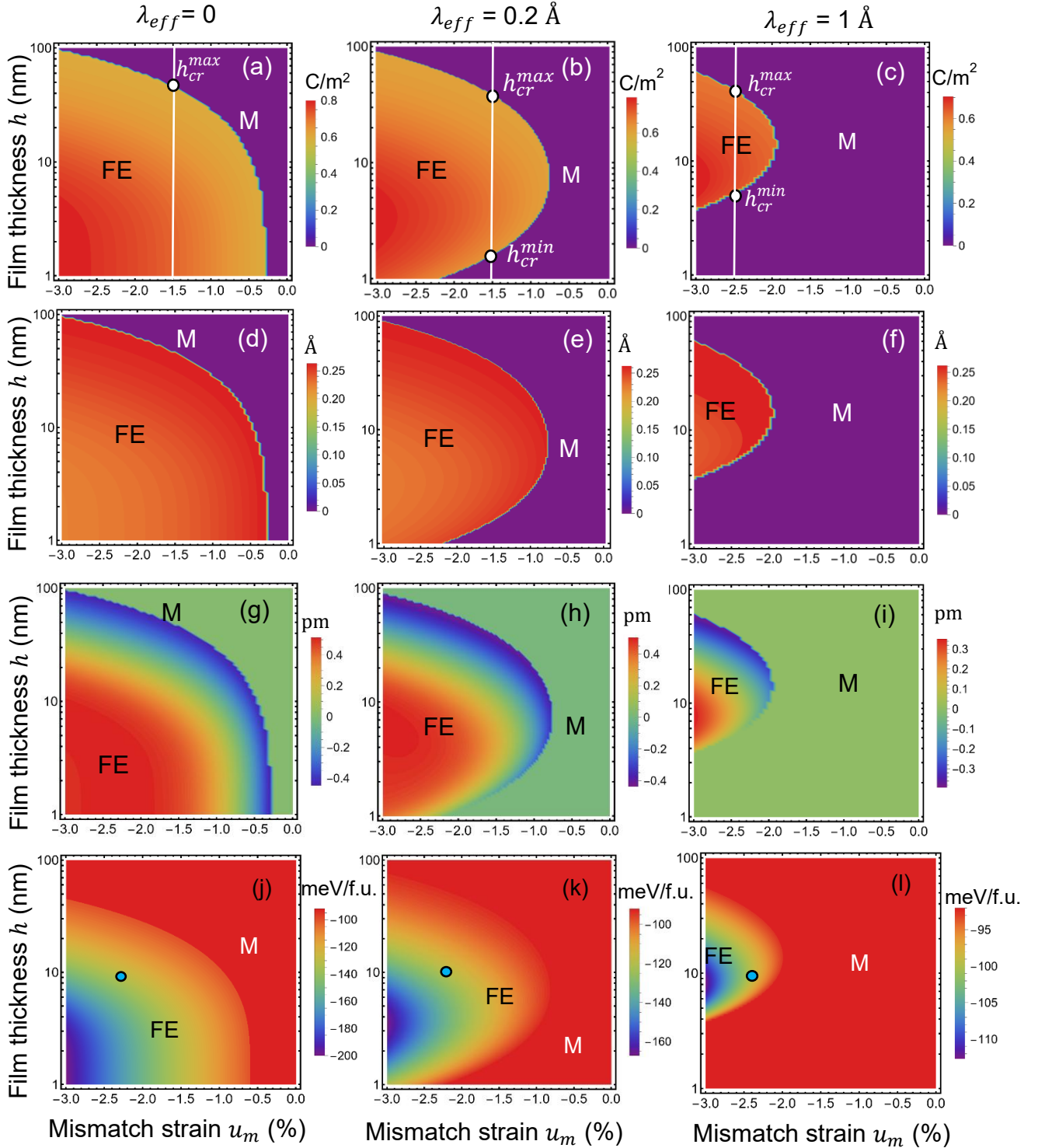

**FIGURE 2.** The absolute value of polarization $\bar{P}_S$ **(a, b, c)**, the amplitude of nonpolar order parameter $|\bar{Q}_{Y2s}|$ **(d, e, f)**, the difference $|\bar{Q}_{Y4s}| - |\bar{Q}_{Y2s}|$ **(g, h, i)**, and the deepest minimum of the free energy density $f_{o-phase}^{min}$ **(j, k, l)** as a function of mismatch strain $u_m$ and HfO$_2$ film thickness $h$ calculated for different

screening lengths $\lambda_{eff} = 0$ **(a, d, g, j)**, 0.2 Å **(b, e, h, k)** and 1 Å **(c, f, i, l)**. Abbreviation "FE" is the FE o-phase, "M" denotes the nonpolar m-phase. White circles in the plots **(a)-(c)** correspond to $h_{cr}^{min}$ and $h_{cr}^{max}$ at fixed $u_m$. Other parameters are $\Lambda_P = 10$ μm and $h_d = 10$ nm.

Color maps of the free energy density of epitaxial HfO$_2$ films as a function of $\bar{P}_3$ and $\bar{Q}_Y$ are shown in **Figs. 3(a) – 3(c)**, where the amplitude $\bar{Q}_Y = Q_{Y2} = Q_{Y4}$ in the upper half-plane and $\bar{Q}_Y = Q_{Y2} = -Q_{Y4}$ in the bottom half-plane. The plots are calculated for the same values of $\lambda_{eff}$ as in **Fig. 2** and fixed other parameters ($h = 10$ nm and $u_m = -2.2$ %). The signs of spontaneous polarization, antipolar and nonpolar order parameters are inter-dependent in the minimum of the free energy due to the trilinear coupling term $\sim \overline{Q_{Y2}Q_{Y4}P_3}$. From **Fig. 3**, the triple product $\tilde{\gamma}_f \overline{Q_{Y2}Q_{Y4}P_3}$ should be negative in the absolute minimum of the free energy (7), being consistent with equations of state (8). It is seen from the dashed curves in **Figs. 3(a) – 3(c)**, that the switching path between the $-\bar{P}_s$ and $+\bar{P}_s$ states goes through the virtual *Ccce* phase, because the height $b_{af}$ of the lowest barrier of polarization switching, which changes from +48 meV/f.u. (at $\lambda_{eff} = 0$ Å) to +49 meV/f.u. (at $\lambda_{eff} = 1$ Å), well corresponds to the energy density of the *Ccce* phase counted from the t-phase (see **Table S3** in Ref. [42]). An activation field $E_{af}$ of polarization reversal, which corresponds to the domain nucleation onset, can be estimated as $E_{af} \cong b_{af}/\bar{P}_s$, that gives 0.72 MV/cm at $\lambda_{eff} = 0$ Å and 0.84 MV/cm at $\lambda_{eff} = 1$ Å at $V_{f.u.} \approx 134$ Å$^3$. The calculated nucleation fields are lower (but of the same order) than the coercive field $E_c \cong 1.05 - 1.35$ MV/cm observed in 10-nm thick HfO$_2$ films [5, 52], as it should be expected from the nucleation rate theory. From our calculations, the barrier of polarization switching and the activation field very weakly increases with increase in $\lambda_{eff}$. At the same time the depth of the deepest potential well decreases with increase in $\lambda_{eff}$ slightly more strongly, changing from 110 meV/f.u. (at $\lambda_{eff} = 0$ Å) to +95 meV/f.u. (at $\lambda_{eff} = 1$ Å). The width of the deepest potential well, estimated using the energy of the bulk m-phase ($f_m = -92$ meV/f.u. counted from the t-phase), decreases with increase in $\lambda_{eff}$ more strongly (see white elliptic contours in **Figs. 3(a) – 3(c)**).

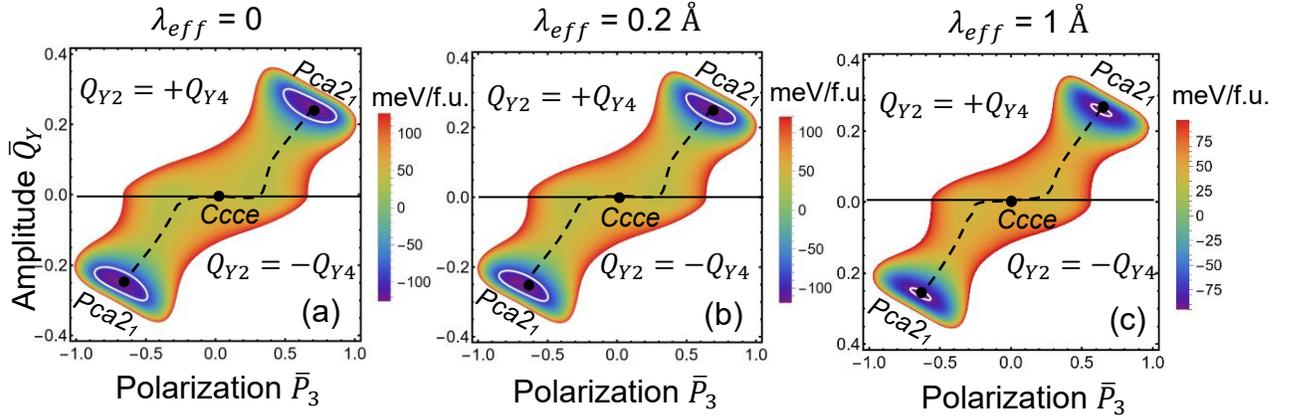

**FIGURE 3.** Free energy density as a function of $\bar{P}_3$ and $\bar{Q}_Y$. The amplitude $\bar{Q}_Y = Q_{Y2} = Q_{Y4}$ in the upper half-plane and $\bar{Q}_Y = Q_{Y2} = -Q_{Y4}$ in the bottom half-plane. The plots are calculated for different screening lengths $\lambda_{eff} = 0$ **(a)**, 0.2 Å **(b)**, 1 Å **(c)**, $h = 10$ nm and $u_m = -2.2$ %. Chosen values of $h$, $u_m$, and $\lambda_{eff}$ correspond to the blue circles in **Figs. 2(j), (k)** and **(l),** respectively. White elliptic contours in the plots correspond to the energy density of the bulk HfO$_2$ m-phase $f_m = -92$ meV/f.u. Dashed curves show the polarization switching path, which corresponds to the lowest energy barrier $f_{Ccce} = +48$ meV/f.u. (counted from the t-phase). Other parameters are the same as in **Fig. 2**.

### C. Epitaxial Nano-Islands of Hafnia-Zirconia

Next let us consider a single-domain HZO nano-island epitaxially grown on a rigid substrate (corresponding geometry is shown in **Fig. 1(c)**). The lateral size $2R$ of the nano-island can be larger or smaller than its height $h$. The assumption $2R \gg h$ allows us to neglect the edge effects in the first approximation, and the elastic strain of the nano-island is almost the same as for the epitaxial film in the case. In the opposite case, $2R \leq h$, a pronounced size effect related with the relaxation of elastic stresses appears due to the edge effects. The relaxation of mismatch strains is proportional to the ratio $\frac{h}{2R}$ at $2R \gg h$ (see Refs. [53, 54]). This allows us to introduce the effective mismatch strain $u_m$:

$$u_m^* = u_m \left(1 - \exp\left[-\frac{h_d}{h}\right]\right) \exp\left[-\frac{h}{2R}\right]. \qquad (13)$$

Hereinafter we regard that the evolution of misfit dislocations inside the island obeys the same low as in epitaxial thin films.

The depolarization field inside the single-domain nano-island of quasi-cylindrical shape can be estimated as [50]:

$$E_3(z) \approx \frac{\bar{P}_3-P_3(z)}{\varepsilon_0\varepsilon_b[1+(h/2R)^2]} - \frac{\bar{P}_3}{\varepsilon_0\varepsilon_b[1+(h/2R)^2]} \frac{\lambda_{eff}}{\lambda_{eff}+h/\varepsilon_b}, \quad (14)$$

where $P_3(z)$ is the out-of-plane polarization directed along z-axis, $\bar{P}_3$ is the z-averaged polarization; $\lambda_{eff}$ is the effective screening length, which contributions come from to the top ionic-electronic screening charge layer and the bottom electrode (see in **Fig. 1(c)**). Note that Eq.(14) for nano-islands transforms into Eq.(6) for thin films in the limiting case $R \to \infty$.

Expressions (7a)-(7f) for renormalized coefficients are approximately valid for nano-islands, at that $u_m^*$ is given by Eq.(13) and depolarization field is given by Eq.(14). The latter leads to the depolarization field contribution $\frac{\lambda_{eff}}{2\varepsilon_0\varepsilon_b(h+\lambda_{eff})[1+(h/2R)^2]}$ in Eq.(7b).

To derive relatively simple equation for the critical height of HZO nano-islands, we should assume that the reversible spontaneous polarization $\bar{P}_s$ relatively weakly depends on the mismatch strain and the island height in the "deep" FE o-phase. Using these assumptions, an approximate transcendental equation for the critical thickness $h_{cr}$ acquires the form:

$$\tilde{\beta}_{\Gamma 3} + \frac{\lambda_{eff}}{2\varepsilon_0\varepsilon_b(h_{cr}+\lambda_{eff})[1+(h_{cr}/2R)^2]} + \frac{g}{h_{cr}\Lambda_P+h_{cr}^2/\pi} - u_m\left(1-\exp\left[-\frac{h_d}{h_{cr}}\right]\right)\exp\left[-\frac{h_{cr}}{2R}\right]\left(\tilde{q}_{13} + \tilde{q}_{23} - \frac{c_{13}+c_{23}}{c_{33}}\tilde{q}_{33}\right) = -\Delta_s, \quad (15)$$

where the positive parameter $\Delta_s$ is introduced in Eq.(12). Considering the same assumptions as for thin HZO films, the parameter $\Delta_s$ can be regarded as mismatch independent. All figures below are calculated using exact Eqs.(7)-(8), because the accuracy of approximate Eq.(15) is not high enough near the boundaries between different phases. However, Eq.(15) well describes qualitatively the critical sizes of HZO nano-islands.

Phase state of the epitaxial HfO$_2$ nano-islands as a function of mismatch strain $u_m$ and their height $h$ is shown in **Fig. 4** for several aspect ratios $h/R = 0.4$ (that corresponds to the strongly flattened island), $h/R = 1$ (that corresponds to the slightly flattened island) and $h/R = 2$ (that corresponds to the island with a square cross-section in the plane, normal to the substrate surface) and fixed effective screening length $\lambda_{eff} = 0.2$ Å. Since the domain formation starts $\lambda_{eff}$ larger than 1 Å, we postpone the case of large screening lengths for next studies.

It is seen that the FE o-phase can be stable at compressive mismatch strains $u_m < u_{min}$ only, where the minimal strain $u_{min}$ changes from $-1.15$ % at $h/R = 0.4$ to $-1.6$ % at $h/R = 2$ due to the size effect of the depolarization field (see Eq.(14)). The area of the FE o-phase region decreases strongly with increase in the aspect ratio $h/R$, being replaced by the reentrant nonpolar m-phase (compare the first, the second and the third columns in **Fig. 4**). The FE o-phase disappears completely for $h > h_{max}$, where $h_{max}$ increases and saturates with increase in $R$, increases with increase in the maximal compressive strain $u_{max}$ and decreases strongly with increase in $\lambda_{eff}$. In particular, the FE o-phase disappears completely for $h > 4\,R$, $\lambda_{eff} = 0.2$ Å and $u_{max} = -3$ %.

Color scale in **Figs. 4(a)-4(c)** is the absolute value of $\bar{P}_s$ in the deepest potential well of the free energy (7). As in the case of thin films, the boundary between the FE o-phase (with $\bar{P}_s > 0$) and nonpolar m-phase (with $\bar{P}_s = 0$) is sharp for nano-islands, being the first order phase transition curve describing the dependence the island critical height $h_{cr}$ on the mismatch strain $u_m$. It is seen from Eq.(15) that the strain-dependent critical height of the ferroelectricity appearance ($h_{cr}^{max}$) is determined by the size dependence of the effective mismatch strain $u_m^*$ (proportional to the function $u_m \left(1 - \exp\left[-\frac{h_d}{h}\right]\right) \exp\left[-\frac{h}{2R}\right]$ according to Eq.(13)). Due to a possible dislocation appearance, the power decay of mismatch strains occurs for $h > h_d$; the lateral relaxation of strains occurs at $h > 2R$. Thus, the estimations $h_{cr}^{max} \leq 10 h_d$ and $h_{cr}^{max} \leq 3R$ may be valid. The critical height of the spontaneous polarization disappearance ($h_{cr}^{min}$) is determined by the size dependence of the depolarization field (proportional to the function $\frac{\lambda_{eff}}{2\varepsilon_0 \varepsilon_b (h + \lambda_{eff})[1 + (h/2R)^2]}$ according to Eq.(14)) and correlation effects (proportional to the function $\frac{g}{h \Lambda_P + h^2/\pi}$). Hence, the estimations $h_{cr}^{min} \gg \lambda_{eff}$ and $h_{cr}^{min} \gg \sqrt{g \varepsilon_0 \varepsilon_b}$, made for thin films, are valid for nano-islands also. The critical heights $h_{cr}^{min}$ and $h_{cr}^{max}$ for the fixed mismatch strain $u_m$ are shown by white circles in **Figs. 4(a)-4(c)**. For incomplete screening conditions ($\lambda_{eff} > 0$), the FE o-phase is thermodynamically stable in the HZO nano-islands with thickness $h_{cr}^{min} < h < h_{cr}^{max}$ when $R > R_{min}$. The minimal lateral size $R_{min}$ is determined by the maximal mismatch strain $u_{max}$. The thicknesses $h_{cr}^{min}$ and $h_{cr}^{max}$ become equal at $R \to R_{min}$ and disappear at $R < R_{min}$.

Color scale in **Figs. 4(d)-4(f)** is the absolute value of $|\bar{Q}_{Y2s}|$ in the deepest potential well. The difference $|\bar{Q}_{Y4s}| - |\bar{Q}_{Y2s}|$ is shown in **Figs. 4(g)-4(i)**. It is seen that the difference $|\bar{Q}_{Y4s}| - |\bar{Q}_{Y2s}|$ is very small: it does not exceed 0.42 pm, that is 60 times smaller than the maximal value of $\bar{Q}_{Y2s}$ (~25 Å). The deepest minimum of the free energy density, $f_{o-phase}^{min}$, is shown in **Figs. 4(j)-4(l)** as a function of $u_m$ and $h$. A red background in the figure corresponds to the bulk m-phase with the energy density $f_m = -92$ meV/f.u. [9].

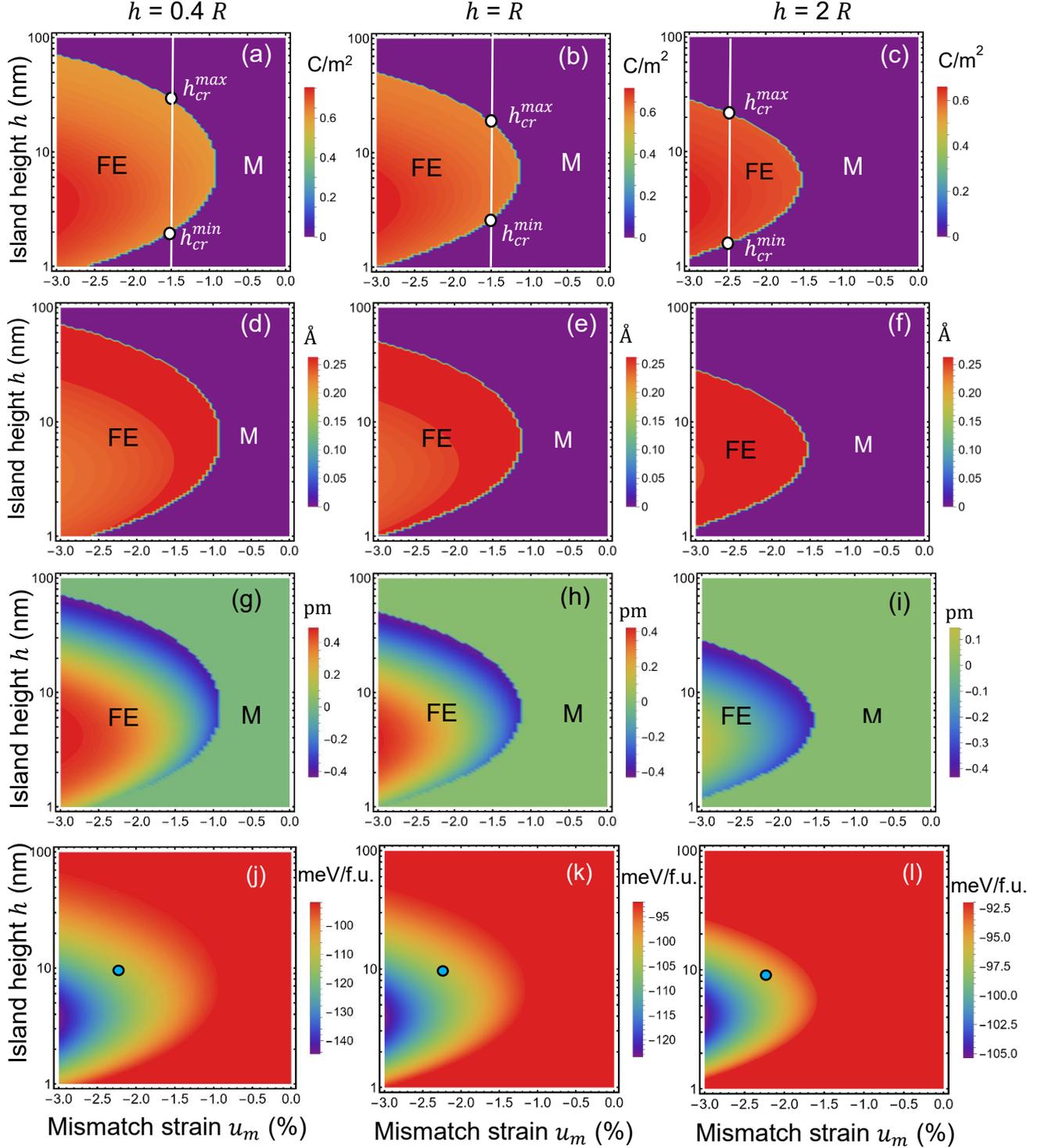

**FIGURE 4.** The absolute value of polarization $\bar{P}_S$ **(a, b, c)**, the amplitude of nonpolar order parameter $|\bar{Q}_{Y2S}|$ **(d, e, f)**, the difference $|\bar{Q}_{Y4S}| - |\bar{Q}_{Y2S}|$ **(g, h, i)** and the deepest minimum of the free energy density $f_{o-phase}^{min}$ **(j, k, l)** as a function of mismatch strain $u_m$ and HfO$_2$ island height $h$ calculated for different aspect ratios $h/R = 0.4$ **(a, d, g, j)**, 1 **(b, e, h, k)** and 2 **(c, f, i, l)**. Abbreviation "FE" is the FE o-phase, "M"

denotes the nonpolar m-phase. White circles in the plots **(a)-(c)** correspond to $h_{cr}^{min}$ and $h_{cr}^{max}$ at fixed $u_m$. Other parameters are $\lambda_{eff} = 0.2$ Å, $\Lambda_P = 10$ µm and $h_d = 10$ nm.

Color maps of the free energy density of epitaxial HfO$_2$ nano-islands as a function of $\bar{P}_3$ and $\bar{Q}_Y$ are shown in **Figs. 5(a) – 5(c)**, where the amplitude $\bar{Q}_Y = Q_{Y2} = Q_{Y4}$ in the upper half-plane and $\bar{Q}_Y = Q_{Y2} = -Q_{Y4}$ in the bottom half-plane. The figures are calculated for the same ratios $h/R$ as in **Fig. 4** and fixed other parameters ($h = 10$ nm, $\lambda_{eff} = 0.2$ Å and $u_m = -2.2$ %). Similarly to the situation in thin HfO$_2$ films, the switching path between the $-\bar{P}_s$ and $+\bar{P}_s$ states goes through the virtual *Ccce* phase in HfO$_2$ nano-islands, because the lowest barrier of polarization switching $b_{af}$ is about +48 meV/f.u., which corresponds to the energy of *Ccce* phase counted from the t-phase (see the color scale and dashed curves in **Figs. 5(a) – 5(c)**). An activation field $E_{af}$ of polarization reversal, estimated as $E_{af} \cong b_{af}/\bar{P}_s$, is about $0.7 - 0.8$ MV/cm at $V_{f.u.} \approx 134$ Å$^3$. The values of $E_{af}$ are very close to the ones calculated in thin films in section B.

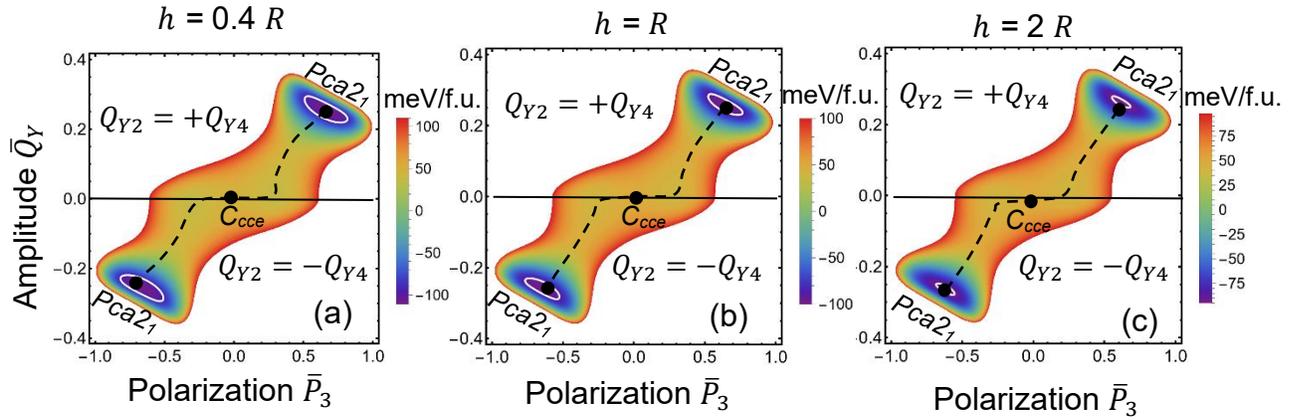

**FIGURE 5.** Free energy density as a function of $\bar{P}_3$ and $\bar{Q}_Y$. The amplitude $\bar{Q}_Y = Q_{Y2} = Q_{Y4}$ in the upper half-plane and $\bar{Q}_Y = Q_{Y2} = -Q_{Y4}$ in the bottom half-plane. The plots are calculated for different radii $R = 25$ nm **(a)**, 10 nm **(b)**, 5 nm **(c)**, $h = 10$ nm, $\lambda_{eff} = 0.2$ Å and $u_m = -2.2$ %. Chosen values of $h$, $u_m$, and $\lambda_{eff}$ correspond to the blue circles in plots **(a)**, **(b)** and **(c)**, respectively. White elliptic contours in the plots correspond to the energy of the bulk m-phase $f_m = -92$ meV/f.u. Dashed curves show the polarization switching path, which corresponds to the lowest energy barrier $f_{Ccce} = +48$ meV/f.u. Other parameters are the same as in **Fig. 4**.

Let us underline an evident similarity of the free energy maps of HfO$_2$ nano-islands, shown in **Fig. 5**, to the free energy maps of HfO$_2$ thin films, shown in **Fig. 3**. Both figures are almost the same, and the increase in $\lambda_{eff}$ leads to the same changes of the energy surface as the increase in $h$. The result can be explained by the similar influence of these parameters on the depolarization field factor of thin films and nano-islands, namely $\frac{\lambda_{eff}}{2\varepsilon_0\varepsilon_b(h+\lambda_{eff})}$ and $\frac{\lambda_{eff}}{2\varepsilon_0\varepsilon_b(h+\lambda_{eff})[1+(h/2R)^2]}$, which increase proportional to $\frac{1}{1+\left(\frac{h}{\lambda_{eff}}\right)}$ with increase in $\lambda_{eff}$. The factor of nano-islands increases proportional to $\frac{1}{1+(h/2R)^2}$ with increase in $h$. Thus, the key factor influencing the free energy surface, phase diagrams and polar properties are the screening length-thickness ratio $\frac{\lambda_{eff}}{h}$ for thin films and the aspect ratio $h/2R$ for nano-islands (at fixed $\lambda_{eff} > 0$).

### D. Applicability of theoretical results to real HZO films

The applicability of the results obtained in sections B and C to real structures, which contain thin or ultra-thin HZO films, is determined by the film growth conditions and the growth mechanism realized in concrete cases. As a rule, three modes of epitaxial film growth are usually distinguished (see, e.g., Ref. [55] and refs therein). In the Frank-van der Merwe regime, atomic monolayers are sequentially formed on the substrate. In the Volmer-Weber regime, nano-islands grow on the substrate, which surface remains partially uncovered by the film material. In the intermediate Stranski-Krastanov regime, a monolayer is formed on the substrate at first (as in the Frank-van der Merwe regime), but after that the Volmer-Weber regime of nano-island growth is realized. The growth regime, that is realized in each specific case, depends on the ratio of surface energies of the film, substrate, and film-substrate interface.

In fact, section B considers the film grown in the Frank–van der Merwe regime, and section C considers the film grown in the Volmer–Weber regime within an idealized approximation of the nano-islands by a cylindrical shape, which allows obtaining semi-analytical results. Real nano-islands can have a wide variety of shapes: domes, pyramids, etc., (see, e.g., [56, 57, 58, 59]). In both cases, the mismatch-dependent critical thicknesses, $h_{cr}^{min}$ and $h_{cr}^{max}$, exist under incomplete screening conditions at $\lambda_{eff} > 0$. The FE o-phase is thermodynamically stable in HZO films with thickness $h_{cr}^{min} < h < h_{cr}^{max}$.

However, there is a significant difference. The critical thicknesses of the film grown in the Frank-van der Merwe regime are determined by the film and substrate lattice constants mismatch $u_m$ (via the epitaxial strains caused by the mismatch) and effective screening length $\lambda_{eff}$ (via the conductivity of electrodes and/or screening conditions at the electrically open surfaces). The critical thickness of the film grown in the Volmer-Weber regime also depends strongly on the average lateral size $2R$ of the nano-island because of the dependence of mismatch strain on the aspect ratio (see Eq.(13), where we neglected the spread of the size in the first approximation). Due to the formation of sufficiently large islands in the HZO films, grown in the Stransky-Krastanov regime, the presence of an additional monolayer at the substrate changes slightly the overall physical picture. Thus, it can be expected that the conclusions obtained for the Volmer-Weber regime are also applicable (at least qualitatively) to the Stransky-Krastanov regime.

Since the modern technological processes for growing thin HZO films implement the Volmer-Weber regime (see, e.g., Refs. [52, 60, 61]), the critical film thicknesses of the ferroelectricity appearance and disappearance give us (at least by an order of magnitude) the diameter $2R$ of the nano-islands, from which the film is composed. At that, the key factors influencing the free energy surface, phase diagrams and corresponding polar properties of nano-sized HZO are screening length – thickness ratio $\lambda_{eff}/h$ (for epitaxial thin films) and the aspect ratio $h/2R$ (for nano-islands). Schematic evolution of the phase diagrams of nanosized HZO subjected to epitaxial mismatch strain $u_m$, which happens with increase in the screening length - thickness ratio $\lambda_{eff}/h$ and/or the aspect ratio $h/2R$, is shown in **Fig. 6**.

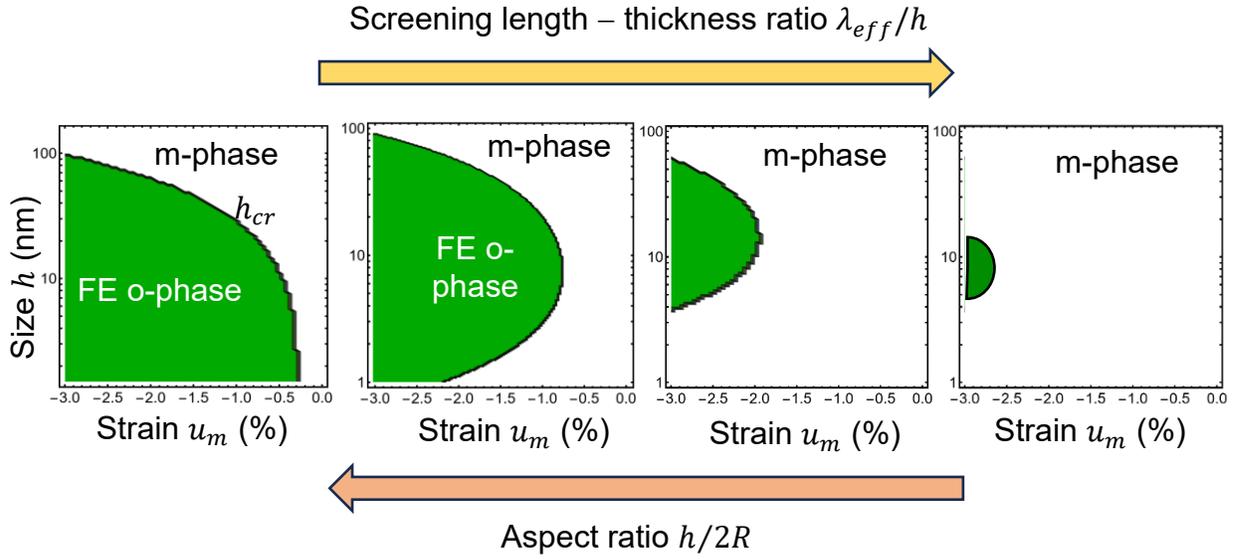

**FIGURE 6.** Schematic evolution of the phase diagrams of nanosized HZO subjected to epitaxial mismatch strain $u_m$, which happens with increase in the screening length – thickness ratio $\lambda_{eff}/h$ and/or with decrease in the aspect ratio $h/2R$. The black solid curve, which separates the FE o-phase from the m-phase, gives the dependence of the critical size $h_{cr}$ on the strain $u_m$ at fixed other parameters.

## III. CONCLUSIONS

Using the LGD free energy functional considering higher powers, biquadratic and trilinear couplings of the polar $\Gamma_{3-}$, nonpolar $Y_{2+}$ and antipolar $Y_{4-}$ modes, we calculated the strain-dependent critical sizes of the ferroelectricity appearance and disappearance, analyzed the mismatch strains influence and size effect of the phase diagrams, polar, antipolar and nonpolar order parameters, and polarization switching barrier in epitaxial $HfO_2$ thin films and nano-islands with the out-of-plane spontaneous polarization.

We have shown that the critical thickness/height of the out-of-plane spontaneous polarization disappearance $h_{cr}^{min}$ is determined by the size dependence of the depolarization field and correlation effects. The critical thickness/height of the ferroelectricity appearance $h_{cr}^{max}$ is determined by the size dependence of the effective mismatch strain $u_m^*$ considering possible appearance of misfit dislocations and lateral relaxion of strains. The key factors influencing the free energy surface, phase diagrams and order parameters of nanosized HZO are thickness-screening length ratio $\frac{\lambda_{eff}}{h}$ (for epitaxial thin films covered by electrodes with effective screening length $\lambda_{eff}$) and the aspect ratio $\frac{h}{2R}$ (for nano-islands with the lateral size $2R$).

In this work we consider the single-domain state of the out-of-plane spontaneous polarization in epitaxial HfO$_2$ thin films and nano-islands, which is stable at small effective screening lengths, $\lambda_{eff} \leq 1$ Å. The domain formation should occur at larger $\lambda_{eff}$; it may decrease the critical thickness of ferroelectricity disappearance, and the case of larger $\lambda_{eff}$ deserves further studies. Our approach can be generalized for HZO thin films and nanoparticles, providing that corresponding parameters of the free energy are known from the first principles calculations.

**Acknowledgements.** The work of A.N.M. is funded by the National Research Foundation of Ukraine (project "Manyfold-degenerated metastable states of spontaneous polarization in nanoferroics: theory, experiment and perspectives for digital nanoelectronics", grant N 2023.03/0132). The work of E.A.E. is funded by the National Research Foundation of Ukraine (project "Silicon-compatible ferroelectric nanocomposites for electronics and sensors", grant N 2023.03/0127). The work of S.V.K. is supported by S.V.K. start-up funds. The work of M.V.S. is funded by the Ministry of Education and Science of Ukraine trough the institutional funding for R&D at Taras Shevchenko National University of Kyiv. Numerical results presented in the work are obtained and visualized using a specialized software, Mathematica 14.0 [62].

**Authors' contribution.** A.N.M. generated the idea of research, formulated the physical problem, performed analytical calculations of the critical size, analyzed results and wrote the manuscript draft. E.A.E. performed symmetry analysis, numerical calculations and prepared corresponding figures. S.V.K. and M.V.S. proposed the concept and worked on improvement of the manuscript. All co-authors discussed the results.

## Supplementary Materials
### APPENDIX S1. LGD-type free energy derived from DFT calculations

Following Delodovici et al. [63] we consider the nonpolar t-phase as the reference (aristo-phase) and the free energy expansion with respect to the powers of the order parameters $Q_{\Gamma 3}$, $Q_{Y2}$ and $Q_{Y4}$, which are introduced as the amplitudes of the corresponding phonon modes normalized to their equilibrium values. When constructing the free energy, one should consider the transformation properties of the order parameters and the symmetry of aristo-phase. The expression for the fee energy with fourth maximal power (having the minimal number of expansion coefficients needed to describe the transition) is:

$$\Delta F_{2-4} = \beta_{200}Q_{\Gamma 3}^2 + \beta_{020}Q_{Y2}^2 + \beta_{002}Q_{Y4}^2 + \gamma_{111}Q_{\Gamma 3}Q_{Y2}Q_{Y4} + \delta_{400}Q_{\Gamma 3}^4 + \delta_{040}Q_{Y2}^4 + \delta_{004}Q_{Y4}^4 +$$
$$\delta_{220}Q_{\Gamma 3}^2 Q_{Y2}^2 + \delta_{202}Q_{\Gamma 3}^2 Q_{Y3}^2 + \delta_{022}Q_{Y3}^2 Q_{Y4}^2 \quad (S1.1)$$

Delodovici et al. [9] have found that the soft mode $\Gamma_{3-}$ is weakly unstable, whereas other distortions, $Y_{2+}$ and $Y_{4-}$, are "hard" modes. So that only the strong trilinear coupling of these three modes, which energy $f_{tr}$ is proportional to the product $Q_{\Gamma 3}Q_{Y2}Q_{Y4}$, can stabilize the FE o-phase. Note the trilinear coupling term $\gamma_{111}Q_{\Gamma 3}Q_{Y2}Q_{Y4}$ alongside the basic biquadratic coupling.

In many cases (for instance, for the first order phase transition) one should include the higher order terms (up to the sixth power):

$$\Delta F_{2-4-6} = \Delta F_{2-4} + (\epsilon_{311}Q_{\Gamma 3}^2 + \epsilon_{131}Q_{Y2}^2 + \epsilon_{113}Q_{Y4}^2)Q_{\Gamma 3}Q_{Y2}Q_{Y4} + \eta_{600}Q_{\Gamma 3}^6 + \eta_{060}Q_{Y2}^6 +$$
$$\eta_{006}Q_{Y4}^6 + \eta_{420}Q_{\Gamma 3}^4 Q_{Y2}^2 + \eta_{402}Q_{\Gamma 3}^4 Q_{Y3}^2 + \eta_{042}Q_{Y3}^4 Q_{Y4}^2 + \eta_{240}Q_{\Gamma 3}^2 Q_{Y2}^4 + \eta_{204}Q_{\Gamma 3}^2 Q_{Y3}^4 +$$
$$\eta_{024}Q_{Y3}^2 Q_{Y4}^4 + \eta_{222}Q_{\Gamma 3}^2 Q_{Y2}^2 Q_{Y4}^2 \quad (S1.2)$$

Using the dependence of the system energy on amplitudes of the $Q_{\Gamma 3}$, $Q_{Y2}$ and $Q_{Y4}$ modes, obtained by Delodovici et al. [9] from DFT, we fitted this dependence with Eqs. (S1.1) and (S1.2) by means of the least squire method under the condition that corresponding free energy has a minimum at $Q_{\Gamma 3} = 1$, $Q_{Y2} = 1$ and $Q_{Y4} = 1$. However, it appears that the model "2-4" (presented by Eq.(S1.1)) gives only qualitative picture of the free energy relief with diminished barrier between the stable phases, while the model "2-6" gives the barriers height close to ab initio results, but the fitting with (S1.2) gives negative values of higher order terms (see **Table S1**). The latter means that the free energy (S1.2) cannot describe the stable phase with the finite values of order parameters in this case.

The way to overcome this problem is to consider the higher power terms (seventh and eighth powers) with constraints based on the stability conditions (like $\xi_{800} > 0$, $\xi_{080} > 0$, $\xi_{008} > 0$, $\xi_{800} + \xi_{080} + \xi_{620} + \xi_{260} + \xi_{440} > 0$ etc.):

$$\Delta F_{2-4-6-8} = \Delta F_{2-4-6} + (\zeta_{511}Q_{\Gamma 3}^4 + \zeta_{151}Q_{Y2}^4 + \zeta_{115}Q_{Y4}^4 + \zeta_{331}Q_{\Gamma 3}^2 Q_{Y2}^2 + \zeta_{313}Q_{\Gamma 3}^2 Q_{Y4}^2 +$$
$$\zeta_{133}Q_{Y2}^2 Q_{Y4}^2)Q_{\Gamma 3}Q_{Y2}Q_{Y4} + \xi_{800}Q_{\Gamma 3}^8 + \xi_{080}Q_{Y2}^8 + \xi_{008}Q_{Y4}^8 + \xi_{620}Q_{\Gamma 3}^6 Q_{Y2}^2 + \xi_{260}Q_{\Gamma 3}^2 Q_{Y2}^6 +$$
$$\xi_{602}Q_{\Gamma 3}^6 Q_{Y4}^2 + \xi_{206}Q_{\Gamma 3}^2 Q_{Y4}^6 + \xi_{062}Q_{Y2}^6 Q_{Y4}^2 + \xi_{026}Q_{Y2}^2 Q_{Y4}^6 + \xi_{440}Q_{\Gamma 3}^4 Q_{Y2}^4 + \xi_{404}Q_{\Gamma 3}^4 Q_{Y4}^4 +$$
$$\xi_{044}Q_{Y2}^4 Q_{Y4}^4 + \xi_{422}Q_{\Gamma 3}^4 Q_{Y2}^2 Q_{Y4}^2 + \xi_{242}Q_{\Gamma 3}^2 Q_{Y2}^4 Q_{Y4}^2 + \xi_{224}Q_{\Gamma 3}^2 Q_{Y2}^2 Q_{Y4}^4 \quad (S1.3)$$

Values of the LGD parameters are summarized in **Tables S1** and **S2**.

**Table S1.** LGD models expansion coefficients for three different models (in eV/f.u.)

| coefficient | model "2-4" | model 2-4-6 | model 2-4-6-8 |
|---|---|---|---|
| $\beta_{200}$ | −0.0343 | −0.0459 | −0.0473 |
| $\beta_{020}$ | 0.0117 | 0.0052 | 0.00455 |
| $\beta_{002}$ | 0.189 | 0.187 | 0.180 |
| $\gamma_{111}$ | −0.836 | −0.980 | −0.508 |
| $\delta_{400}$ | 0.103, | 0.141 | 0.1495 |
| $\delta_{040}$ | 0.092 | 0.114 | 0.1178 |
| $\delta_{004}$ | 0.0161 | 0.0224 | 0.0705 |
| $\delta_{220}$ | 0.136, | 0.434 | 0.473 |
| $\delta_{202}$ | 0.111 | 0.244 | 0.179 |
| $\delta_{022}$ | 0.0860 | 0.329 | 0.373 |
| $\epsilon_{311}$ | 0 | -0.203 | -0.2944 |
| $\epsilon_{131}$ | 0 | -0.426 | -0.3898 |
| $\epsilon_{113}$ | 0 | -0.301 | -0.3432 |
| $\eta_{600}$ | 0 | -0.0280 | -0.0426 |
| $\eta_{060}$ | 0 | -0.0158 | -0.0228 |
| $\eta_{006}$ | 0 | -0.0046 | -0.0894 |
| $\eta_{042}$ | 0 | 0.0369 | -0.0182 |
| $\eta_{024}$ | 0 | 0.027 | -0.0136 |
| $\eta_{402}$ | 0 | 0.0156 | 0.0659 |
| $\eta_{204}$ | 0 | 0.0330 | 0.0845 |
| $\eta_{420}$ | 0 | -0.0413 | -0.1053 |
| $\eta_{240}$ | 0 | 0.0370 | -0.0250 |
| $\eta_{222}$ | 0 | 0.293 | -0.2785 |

**Table S2.** The seventh and eighth powers expansion coefficients for the "2-4-6-8" model (in eV/f.u.).

| $\zeta_{511}$ | $\zeta_{151}$ | $\zeta_{115}$ | $\zeta_{331}$ | $\zeta_{313}$ | $\zeta_{133}$ | $\xi_{062}$ |
|---|---|---|---|---|---|---|
| -0.161 | -0.177 | -0.153 | -0.0973 | -0.0394 | -0.158 | 0.0111 |
| $\xi_{800}$ | $\xi_{080}$ | $\xi_{008}$ | $\xi_{620}$ | $\xi_{260}$ | $\xi_{602}$ | $\xi_{206}$ |
| 0.00763 | 0.00365 | 0.0446 | 0.041 | 0.04023 | 0.0234 | 0.0219 |
| $\xi_{440}$ | $\xi_{404}$ | $\xi_{044}$ | $\xi_{422}$ | $\xi_{242}$ | $\xi_{224}$ | $\xi_{026}$ |
| 0.00693 | -0.08156 | 0.0403 | 0.3153 | 0.32307 | 0.3420 | 0.000395 |

The colors maps of the free energy density as the function of order parameters the order parameters are shown in **Figs. S1** and **S2** for cross-sections near the equilibrium points. It is seen that there are four equilibrium states with the different signs of order parameters ($Q_{\Gamma 3} = \pm 1$, $Q_{Y2} = \mp 1$, $Q_{Y2} = \pm 1$). It should be noted that there are two shallow minima with energy $-0.0039$ eV/f.u. at $Q_{\Gamma 3} \approx \pm 0.41$, $Q_{Y2} = 0$, $Q_{Y2} = 0$ (see **Fig. S2(a)**) corresponding to the ferroelectric phase (Ae2a space group), but its energy is much higher than that of the experimentally observed FE o-phase $Pca2_1$.

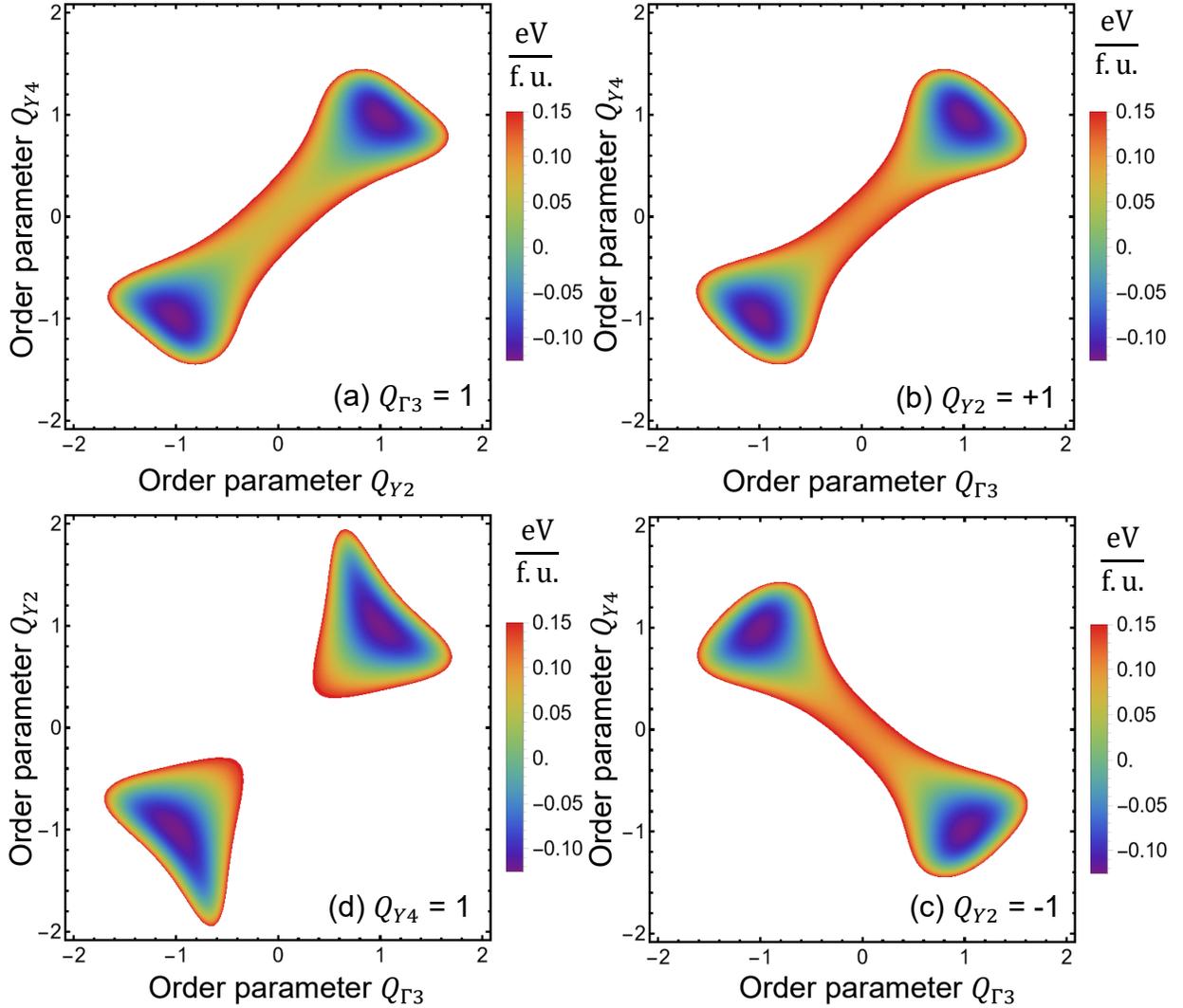

**FIGURE S1.** Color maps showing the dependence of the free energy density (S2.3) on the order parameters for $Q_{\Gamma 3} = 1$ **(a)**, $Q_{Y2} = +1$ **(b)**, $Q_{Y2} = -1$ **(c)**, and $Q_{Y4} = 1$ **(d)**. The LGD parameters for HfO$_2$ are listed in **Tables S1** and **S2**.

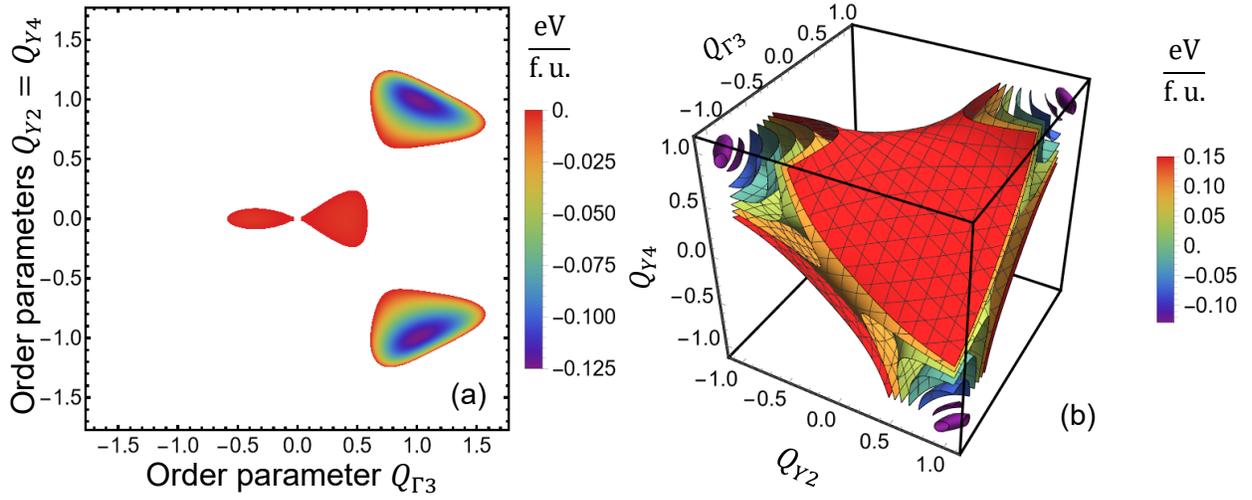

**FIGURE S2. (a)** Color map showing the dependence of the free energy density (S2.3) on the order parameters for specific case $Q_{Y2} = Q_{Y4}$. **(b)** The surfaces of the constant energy in the 3D phase space of order parameters. The LGD parameters for HfO$_2$ are listed in **Tables S1** and **S2**.

To compare the energies of bulk monoclinic phase, which is believed to be the most stable phase of bulk HZO compounds, with nanoscale phase, we listed for comparison the results of Materlik et al. in **Table S3.** Setting the energy scales to the same energy of tetragonal phase (0 meV/f.u.) we estimated the energy of all the phases we are interested in (see the fourth row from **Table S3**).

**Table S3**. Equilibrium energy density values (in meV/f.u.) for different phases, calculated from the first principles

| Source | m-phase | t-phase | Ccce phase | FE o-phase |
|---|---|---|---|---|
| Delodovici et al. [9] | - | -48 | 0 | -126 |
| Materlik et al. [64] | 0 | 92 | -- | 62 |
| Rescaled results of [9] and [64] | -92 | 0 | 48 | -78 |

# APPENDIX S2. Anisotropic misfit strain for [001] oriented films of multiferroic-uniaxial ferroelectric with aristo-phase of mmm symmetry

Assuming that order parameters are homogeneous and neglecting the gradient energy, phenomenological free energy at fixed values of strain could be written as follows:

$$\Delta F = \Delta F_{FE} + \Delta F_{elast} \qquad (S2.1a)$$

$$\Delta F_{FE} = \beta_\Gamma P_3^2 + \delta_\Gamma P_3^4 + \eta_\Gamma P_3^6 + \xi_\Gamma P_3^8 + $$
$$\beta_\Psi \Psi^2 + \delta_\Psi \Psi^4 + \eta_\Psi \Psi^6 + \xi_\Psi \Psi^8 + $$
$$\beta_\Phi \Phi^2 + \delta_\Phi \Phi^4 + \eta_\Phi \Phi^6 + \xi_\Phi \Phi^8 + $$
$$(\gamma + \epsilon_\Gamma P_3^2 + \epsilon_\Psi \Psi^2 + \epsilon_\Phi \Phi^2 + \zeta_{\Gamma\Phi} P_3^2 \Phi^2 + \zeta_{\Gamma\Psi} P_3^2 \Psi^2 + \zeta_{\Phi\Psi} \Phi^2 \Psi^2 + \zeta_\Gamma P_3^4 + \zeta_\Phi \Phi^4$$
$$+ \zeta_\Psi \Psi^4) P_3 \Psi \Phi + \delta_{\Gamma\Psi} P_3^2 \Psi^2 + \delta_{\Gamma\Phi} P_3^2 \Phi^2 + \delta_{\Psi\Phi} \Psi^2 \Phi^2 + \eta_{\Gamma\Psi\Phi} P_3^2 \Psi^2 \Phi^2$$
$$+ \eta_{\Gamma\Psi\Psi} P_3^2 \Psi^4 + \eta_{\Gamma\Gamma\Psi} P_3^4 \Psi^2 + \eta_{\Gamma\Gamma\Phi} P_3^4 \Phi^2 + \xi_{\Gamma\Psi} P_3^4 \Psi^4$$

$$(S2.1b)$$

$$\Delta F_{elast} = -(q_{13}u_1 + q_{23}u_2 + q_{33}u_3)P_3^2 - (z_{133}u_1 + z_{233}u_2 + z_{333}u_3)P_3^4$$
$$- (q_{1\Psi}u_1 + q_{2\Psi}u_2 + q_{3\Psi}u_3)\Psi^2 - (q_{1\Psi\Psi}u_1 + q_{2\Psi\Psi}u_2 + q_{3\Psi\Psi}u_3)\Psi^4$$
$$- (q_{1\Phi}u_1 + q_{2\Phi}u_2 + q_{3\Phi}u_3)\Phi^2 - (r_{13}u_1 + r_{23}u_2 + r_{33}u_3)P_3\Psi\Phi$$
$$+ \frac{1}{2}(c_{11}u_1^2 + c_{22}u_2^2 + c_{33}u_3^2) + (c_{12}u_1u_2 + c_{13}u_1u_3 + c_{23}u_2u_3)$$
$$+ \frac{1}{2}(c_{66}u_6^2 + c_{55}u_5^2 + c_{44}u_4^2)$$

$$(S2.1c)$$

Here we introduced a "physical" order parameter, related with the dimensionless amplitudes of the polar, nonpolar and antipolar modes in the following way

$$P_3 = P_0 Q_{\Gamma_{3-}}, \quad \Psi = \Psi_0 Q_{Y_{2+}}, \quad \Phi = \Phi_0 Q_{Y_{4-}} \qquad (S2.1d)$$

Here $P_0 = 0.55 \, C/m^2$, $\Psi_0 = 0.278 \, Å$, $\Phi_0 = 0.268 \, Å$ are the maximal polarization and the maximum atomic displacements respectively [9]. Voight matrix notations are used in Eq.(S2.1)

$$c_{1111} = c_{11}, \quad c_{1122} = c_{12}, \quad c_{1212} = c_{44}, \text{etc.} \qquad (S2.2a)$$

$$q_{1111} = q_{11}, \quad q_{1122} = q_{12} \qquad (S2.2b)$$

$$u_{11} = u_1, \quad u_{22} = u_2, \quad 2u_{12} = u_6, \text{etc.} \qquad (S2.2c)$$

Note that for LGD expansion coefficients at given strain tensor are used in Eq.(S2.1). Modified Hooke's law could be obtained from the relation $\sigma_{ij} = +\partial(\Delta F_{FE})/\partial u_{ij}$:

$$\sigma_{11} = c_{11}u_1 + c_{12}u_2 + c_{13}u_3 - q_{13}P_3^2 - z_{133}P_3^4 - q_{1\Psi}\Psi^2 - q_{1\Psi\Psi}\Psi^4 - q_{1\Phi}\Phi^2 - r_{13}P_3\Psi\Phi$$
(S2.3a)

$$\sigma_{22} = c_{12}u_1 + c_{22}u_2 + c_{23}u_3 - q_{23}P_3^2 - z_{233}P_3^4 - q_{2\Psi}\Psi^2 - q_{2\Psi\Psi}\Psi^4 - q_{2\Phi}\Phi^2 - r_{23}P_3\Psi\Phi$$
(S2.3b)

$$\sigma_{33} = c_{13}u_1 + c_{23}u_2 + c_{33}u_3 - q_{33}P_3^2 - z_{333}P_3^4 - q_{3\Psi}\Psi^2 - q_{3\Psi\Psi}\Psi^4 - q_{3\Phi}\Phi^2 - r_{33}P_3\Psi\Phi$$
(S2.3c)

$$c_{44}u_4 = \sigma_{23} \quad \text{(S2.3d)}$$
$$c_{55}u_5 = \sigma_{13} \quad \text{(S2.3e)}$$
$$c_{66}u_6 = \sigma_{12} \quad \text{(S2.3f)}$$

The solution for the misfit of thin film with its substrate is well known. For the film with normal along $X_3$ one has the following relations for some of stress and strain components:

$$\sigma_{13} = \sigma_{23} = \sigma_{33} = 0 \quad \text{(S2.4a)}$$
$$u_1 = u_1^{(m)}, \quad u_2 = u_2^{(m)}, \quad u_6 = u_6^{(m)} \quad \text{(S2.4b)}$$

Here $u_1^{(m)}$, $u_2^{(m)}$ and $u_6^{(m)}$ are components of anisotropic misfit strain, two diagonal components are determined by the difference of lattice constants in corresponding direction, while $u_6^{(m)}$ is the difference between the corresponding angles of the unit cells of the film and substrate. Taking (S2.3) and (S2.4) into account

$$\sigma_{11} = c_{11}u_1^{(m)} + c_{12}u_2^{(m)} + c_{13}u_3 - q_{13}P_3^2 - z_{133}P_3^4 - q_{1\Psi}\Psi^2 - q_{1\Psi\Psi}\Psi^4 - q_{1\Phi}\Phi^2 - r_{13}P_3\Psi\Phi$$
(S2.5a)

$$\sigma_{22} = c_{12}u_1^{(m)} + c_{22}u_2^{(m)} + c_{23}u_3 - q_{23}P_3^2 - z_{233}P_3^4 - q_{2\Psi}\Psi^2 - q_{2\Psi\Psi}\Psi^4 - q_{2\Phi}\Phi^2$$
$$- r_{23}P_3\Psi\Phi$$
(S2.5b)

$$0 = c_{13}u_1^{(m)} + c_{23}u_2^{(m)} + c_{33}u_3 - q_{33}P_3^2 - z_{333}P_3^4 - q_{3\Psi}\Psi^2 - q_{3\Psi\Psi}\Psi^4 - q_{3\Phi}\Phi^2 - r_{33}P_3\Psi\Phi$$
(S2.5c)

$$u_4 = 0, \; u_5 = 0, \; c_{66}u_6^{(m)} = \sigma_{12}. \quad \text{(S2.5d)}$$

The unknown strain component $u_3$ could be found from (S2.5) as

$$u_3 = -\frac{c_{13}}{c_{33}}u_1^{(m)} - \frac{c_{23}}{c_{33}}u_2^{(m)} + \frac{1}{c_{33}}(q_{33}P_3^2 + z_{333}P_3^4 + q_{3\Psi}\Psi^2 + q_{3\Phi}\Phi^2 + q_{3\Psi\Psi}\Psi^4 + r_{33}P_3\Psi\Phi)$$
(S2.6a)

The equation of state for polarization along the polar axis $x_3$ could be found from the minimization of (S2.1) with respect to $P_3$:

$2\beta_\Gamma P_3 + 4\delta_\Gamma P_3^3 + 6\eta_\Gamma P_3^5 + 8\xi_\Gamma P_3^7 + (\gamma + 3\epsilon_\Gamma P_3^2 + \epsilon_\Psi \Psi^2 + \epsilon_\Phi \Phi^2 + 3\zeta_{\Gamma\Phi} P_3^2 \Phi^2 + 3\zeta_{\Gamma\Psi} P_3^2 \Psi^2 + \zeta_{\Phi\Psi} \Phi^2 \Psi^2 + 5\zeta_\Gamma P_3^4 + \zeta_\Phi \Phi^4 + \zeta_\Psi \Psi^4)\Psi\Phi + 2\delta_{\Gamma\Psi} P_3 \Psi^2 + 2\delta_{\Gamma\Phi} P_3 \Phi^2 + 2\eta_{\Gamma\Psi\Phi} P_3 \Psi^2 \Phi^2 + 2\eta_{\Gamma\Psi\Psi} P_3 \Psi^4 + 4(\eta_{\Gamma\Gamma\Psi} \Psi^2 + \eta_{\Gamma\Gamma\Phi} \Phi^2) P_3^3 + 4\xi_{\Gamma\Psi} P_3^3 \Psi^4 - 2(q_{13} u_1 + q_{23} u_2 + q_{33} u_3) P_3 - 4(z_{133} u_1 + z_{233} u_2 + z_{333} u_3) P_3^3 - (r_{13} u_1 + r_{23} u_2 + r_{33} u_3)\Psi\Phi = 0$

(S2.7a)

After the substitution of (S2.6a) and (S2.4b) into Eq.(S2.7a)

$2\beta_\Gamma P_3 + 4\delta_\Gamma P_3^3 + 6\eta_\Gamma P_3^5 + 8\xi_\Gamma P_3^7 + (\gamma + 3\epsilon_\Gamma P_3^2 + \epsilon_\Psi \Psi^2 + \epsilon_\Phi \Phi^2 + 3\zeta_{\Gamma\Phi} P_3^2 \Phi^2 + 3\zeta_{\Gamma\Psi} P_3^2 \Psi^2 + \zeta_{\Phi\Psi} \Phi^2 \Psi^2 + 5\zeta_\Gamma P_3^4 + \zeta_\Phi \Phi^4 + \zeta_\Psi \Psi^4)\Psi\Phi + 2\delta_{\Gamma\Psi} P_3 \Psi^2 + 2\delta_{\Gamma\Phi} P_3 \Phi^2 + 2\eta_{\Gamma\Psi\Phi} P_3 \Psi^2 \Phi^2 + 2\eta_{\Gamma\Psi\Psi} P_3 \Psi^4 + 4\eta_{\Gamma\Gamma\Psi} P_3^3 \Psi^2 + 4\xi_{\Gamma\Psi} P_3^3 \Psi^4 - u_1^{(m)}[2q_{13} P_3 + 4z_{133} P_3^3 + r_{13}\Psi\Phi] - u_2^{(m)}[2q_{23} P_3 + 4z_{233} P_3^3 + r_{23}\Psi\Phi] - [2q_{33} P_3 + 4z_{333} P_3^3 + r_{33}\Psi\Phi]\left(-\frac{c_{13}}{c_{33}} u_1^{(m)} - \frac{c_{23}}{c_{33}} u_2^{(m)} + \frac{1}{c_{33}}(q_{33} P_3^2 + z_{333} P_3^4 + q_{3\Psi} \Psi^2 + q_{3\Phi} \Phi^2 + q_{3\Psi\Psi} \Psi^4 + r_{33} P_3 \Psi\Phi)\right) = 0$  (S2.7b)

Finally, after the re-grouping of the terms, one could get the following:

$2\left(\beta_\Gamma - u_1^{(m)}\left[q_{13} - \frac{c_{13}}{c_{33}} q_{33}\right] - u_2^{(m)}\left[q_{23} - \frac{c_{23}}{c_{33}} q_{33}\right]\right) P_3 + 4\left(\delta_\Gamma - \frac{q_{33}^2}{2c_{33}} - u_1^{(m)}\left[z_{133} - \frac{c_{13}}{c_{33}} z_{333}\right] - u_{22}^{(m)}\left[z_{233} - \frac{c_{23}}{c_{33}} z_{333}\right]\right) P_3^3 + 6\left(\eta_\Gamma - \frac{q_{33} z_{333}}{c_{33}}\right) P_3^5 + 8\left(\xi_\Gamma - \frac{z_{333}^2}{2c_{33}}\right) P_3^7 + 2\left(\delta_{\Gamma\Psi} - \frac{q_{33} q_{3\Psi}}{c_{33}}\right) P_3 \Psi^2 + 2\left(\delta_{\Gamma\Phi} - \frac{q_{33} q_{3\Phi}}{c_{33}}\right) P_3 \Phi^2 + \left(2\eta_{\Gamma\Psi\Phi} - \frac{r_{33}^2}{c_{33}}\right) P_3 \Psi^2 \Phi^2 + \left(\left[\epsilon_\Psi - \frac{r_{33}}{c_{33}} q_{3\Psi}\right]\Psi^2 + \left[\epsilon_\Phi - \frac{r_{33}}{c_{33}} q_{3\Phi}\right]\Phi^2 + \left[\zeta_\Psi - \frac{r_{33}}{c_{33}} q_{3\Psi\Psi}\right]\Psi^4 + 3\zeta_{\Gamma\Phi} P_3^2 \Phi^2 + 3\zeta_{\Gamma\Psi} P_3^2 \Psi^2 + \zeta_{\Phi\Psi} \Phi^2 \Psi^2 + \zeta_\Phi \Phi^4\right)\Psi\Phi + 3\left(\epsilon_\Gamma - \frac{q_{33} r_{33}}{c_{33}}\right) P_3^2 \Psi\Phi + 5\left(\zeta_\Gamma - \frac{z_{333} r_{33}}{c_{33}}\right) P_3^4 \Psi\Phi + \left(\gamma - u_1^{(m)}\left[r_{13} - \frac{c_{13}}{c_{33}} r_{33}\right] - u_2^{(m)}\left[r_{23} - \frac{c_{23}}{c_{33}} r_{33}\right]\right)\Psi\Phi + 2\left(\eta_{\Gamma\Psi\Psi} - \frac{q_{33} q_{3\Psi\Psi}}{c_{33}}\right) P_3 \Psi^4 + 4\left(\eta_{\Gamma\Gamma\Psi} - \frac{z_{333} q_{3\Psi}}{c_{33}}\right) P_3^3 \Psi^2 + 4\left(\eta_{\Gamma\Gamma\Phi} - \frac{z_{333} q_{3\Phi}}{c_{33}}\right) P_3^3 \Phi^2 + 4\left(\xi_{\Gamma\Psi} - \frac{z_{333} q_{3\Psi\Psi}}{c_{33}}\right) P_3^3 \Psi^4 = 0$  (S2.7c)

Finally, in the case of zero roto-striction and transversally isotropic misfit strain, one could get the following from Eq.(S2.7c):

$$2\left(\beta_\Gamma - u_m\left[q_{13} + q_{23} - \frac{c_{13}+c_{23}}{c_{33}}q_{33}\right]\right)P_3 + 4\left(\delta_\Gamma - \frac{q_{33}^2}{2c_{33}} - u_m\left[z_{133} + z_{233} - \right.\right.$$

$$\left.\left.\frac{c_{13}+c_{23}}{c_{33}}z_{333}\right]\right)P_3^3 + 6\left(\eta_\Gamma - \frac{q_{33}z_{333}}{c_{33}}\right)P_3^5 + 8\left(\xi_\Gamma - \frac{z_{333}^2}{2c_{33}}\right)P_3^7 + 2\delta_{\Gamma\Psi}P_3\Psi^2 + 2\delta_{\Gamma\Phi}P_3\Phi^2 +$$

$$\left(2\eta_{\Gamma\Psi\Phi} - \frac{r_{33}^2}{c_{33}}\right)P_3\Psi^2\Phi^2 + \left(\gamma - u_m\left[r_{13} + r_{23} - \frac{c_{13}+c_{23}}{c_{33}}r_{33}\right]\right)\Psi\Phi + (\epsilon_\Psi\Psi^2 + \epsilon_\Phi\Phi^2 + \zeta_\Psi\Psi^4 +$$

$$3\zeta_{\Gamma\Phi}P_3^2\Phi^2 + 3\zeta_{\Gamma\Psi}P_3^2\Psi^2 + \zeta_{\Phi\Psi}\Phi^2\Psi^2 + \zeta_\Phi\Phi^4)\Psi\Phi + 3\left(\epsilon_\Gamma - \frac{q_{33}r_{33}}{c_{33}}\right)P_3^2\Psi\Phi + 5\left(\zeta_\Gamma - \right.$$

$$\left.\frac{z_{333}r_{33}}{c_{33}}\right)P_3^4\Psi\Phi + 2\eta_{\Gamma\Psi\Psi}P_3\Psi^4 + 4\eta_{\Gamma\Gamma\Psi}P_3^3\Psi^2 + 4\eta_{\Gamma\Gamma\Phi}P_3^3\Phi^2 + 4\xi_{\Gamma\Psi}P_3^3\Psi^4 = 0 \qquad (S2.7d)$$

## APPENDIX S3. Striction constants of hafnia estimated from the first principles

### S3.1. Piezoelectric effect and electrostriction

Piezoelectric effect tensors were calculated for hafnia by Dutta et al. [65]. The well-known expression for the relationship between tensors of piezoelectric $d_{ijk}$ and electrostriction $Q_{jkmn}$ effects is

$$d_{ijk} = 2\varepsilon_0\varepsilon_{im}P_nQ_{jkmn}, \qquad (S3.1)$$

here $\varepsilon_0$ is the universal dielectric constant, $\varepsilon_{im}$ is the tensor of dielectric permittivity and $P_n$ is the spontaneous polarization component. The evident form of Eq.(S3.1) for the case of polar orthorhombic structure is summarized in **Table S4.**

**Table S4**. The phenomenological relations

| Component designation | Tensor form | Voight from |
|---|---|---|
| $d_{311} = d_{31}$ | $2\varepsilon_0\varepsilon_{33}P_3Q_{1133}$ | $2\varepsilon_0\varepsilon_3 P_3 Q_{13}$ |
| $d_{322} = d_{32}$ | $2\varepsilon_0\varepsilon_{33}P_3Q_{2233}$ | $2\varepsilon_0\varepsilon_3 P_3 Q_{23}$ |
| $d_{333} = d_{33}$, | $2\varepsilon_0\varepsilon_{33}P_3Q_{3333}$ | $2\varepsilon_0\varepsilon_3 P_3 Q_{33}$ |
| $d_{113} \stackrel{\text{def}}{=} d_{15}/2$ | $2\varepsilon_0\varepsilon_{11}P_3Q_{1313}$ | $d_{15} = \varepsilon_0\varepsilon_1 P_3 Q_{55}$ |
| $d_{223} \stackrel{\text{def}}{=} d_{24}/2$ | $2\varepsilon_0\varepsilon_{22}P_3Q_{2323}$ | $d_{24} = \varepsilon_0\varepsilon_2 P_3 Q_{44}$ |

Using dielectric permittivity calculated by Zhao and Vanderbilt [66]

$$\varepsilon_{11}=23,\ \varepsilon_{22}=18,\ \varepsilon_{33}=20, \qquad (S3.2)$$

and polarization value, estimated from the first principles

$$P_3 = P_S = 0.55 \text{ C/m}^2, \qquad (S3.3)$$

one could get the following values for the electrostriction coefficients (in m$^4$/C$^2$ units):

$$Q_{13} = -0.0139, Q_{23} = -0.00853, Q_{33} = -0.008423, Q_{55} = -0.0241, Q_{44} = 0.0847. \quad (S3.4)$$

Next, we recall the relation between electrostriction strain and stress coefficients in the form

$$q_{ijkl} = c_{ijmn} Q_{mnkl} \qquad (S3.5a)$$

Or, in the evident form, the case of orthorhombic symmetry

$$q_{13} = c_{11} Q_{13} + c_{12} Q_{23} + c_{13} Q_{33}, \qquad (S3.5b)$$

$$q_{23} = c_{12} Q_{13} + c_{22} Q_{23} + c_{23} Q_{33}, \qquad (S3.5c)$$

$$q_{33} = c_{13} Q_{13} + c_{23} Q_{23} + c_{33} Q_{33}. \qquad (S3.5d)$$

**Table S5.** The elastic stiffness tensor for $HfO_2$, taken from Ref.[65].

| component | $c_{11}$ | $c_{12}$ | $c_{13}$ | $c_{22}$ | $c_{23}$ | $c_{33}$ | $c_{44}$ | $c_{55}$ | $c_{66}$ |
|---|---|---|---|---|---|---|---|---|---|
| value (GPa) | 413.6 | 162.3 | 123.4 | 407.8 | 132.8 | 394.6 | 94.4 | 98.0 | 140.4 |

Using Eqs.(S3.5b)-(S3.5d) and compliance values (see **Table S5**) one could easily get the following values (in m/F units):

$$q_{13} = -8.1800 \times 10^9, q_{23} = -6.8545 \times 10^9, q_{33} = -6.1736 \times 10^9 \qquad (S3.6)$$

### S3.2. Trilinear striction coefficients

Delodovici et al. [9] determined strain dependence for some of expansion coefficients from Eq.(1), considering volume-preserving strain with $u_{ii} = 0$. The full strain tensor is diagonal with the following nonzero components

$$u_{xx} = u_a, \quad u_{yy} = u_{zz} = -u_a/2, \qquad (S3.7a)$$

$$u_{yy} = u_b, \quad u_{xx} = u_{zz} = -u_b/2, \qquad (S3.7b)$$

$$u_{zz} = u_c, \quad u_{yy} = u_{xx} = -u_c/2. \qquad (S3.7c)$$

Here $u_a$, $u_b$ and $u_c$ are strain amplitudes along corresponding axis for three different cases. It is seen that volume-conservation condition is valid.

Let us introduce the strain dependent trilinear coupling coefficients,

$$\gamma_u \stackrel{\text{def}}{=} \gamma - r_{13} u_{xx} - r_{23} u_{yy} - r_{33} u_{zz} \qquad (S3.8)$$

Substitution of (S3.7) into (S3.8) gives the dependencies:

$$\gamma_a = \gamma - \left(r_{13} - \frac{r_{23}+r_{33}}{2}\right) u_a, \gamma_b = \gamma - \left(r_{23} - \frac{r_{13}+r_{33}}{2}\right) u_b, \gamma_c = \gamma - \left(r_{33} - \frac{r_{23}+r_{13}}{2}\right) u_c \qquad (S3.9)$$

Introducing the following designations for the derivative of the trilinear coupling coefficients with respect to strain components:

$$\frac{\partial \gamma_a}{\partial u_a} \stackrel{\text{def}}{=} \gamma_{u_a}, \quad \frac{\partial \gamma_b}{\partial u_b} \stackrel{\text{def}}{=} \gamma_{u_a} \quad \text{and} \quad \frac{\partial \gamma_c}{\partial u_c} \stackrel{\text{def}}{=} \gamma_{u_c}. \tag{S3.10}$$

Considering Eqs.(S3.10) as the system of equation for unknown $r_{13}$, $r_{23}$ and $r_{33}$, its solution could be found as

$$\tilde{r}_{13} = -\frac{2}{3}\gamma_{u_a} + c, \quad \tilde{r}_{23} = -\frac{2}{3}\gamma_{u_b} + c, \quad \tilde{r}_{33} = -\frac{2}{3}\gamma_{u_c} + c. \tag{S3.11}$$

However, since Eqs.(S3.10) are not independent once due to the relation $\gamma_{u_a} + \gamma_{u_b} + \gamma_{u_c} = 0$, only the differences of the striction coefficients can be determined. That is why the striction coefficients depend on the unknown constant $c$. Delodovici et al. [9] obtained the following values

$$\gamma_{u_a} = -1.0, \quad \gamma_{u_b} = 0.18, \quad \gamma_{u_c} = 0.82. \tag{S3.12}$$

Following Delodovici et al. ideology we should assume that $\tilde{r}_{13} + \tilde{r}_{23} + \tilde{r}_{33} = 0$, and so put the constant $c = 0$.

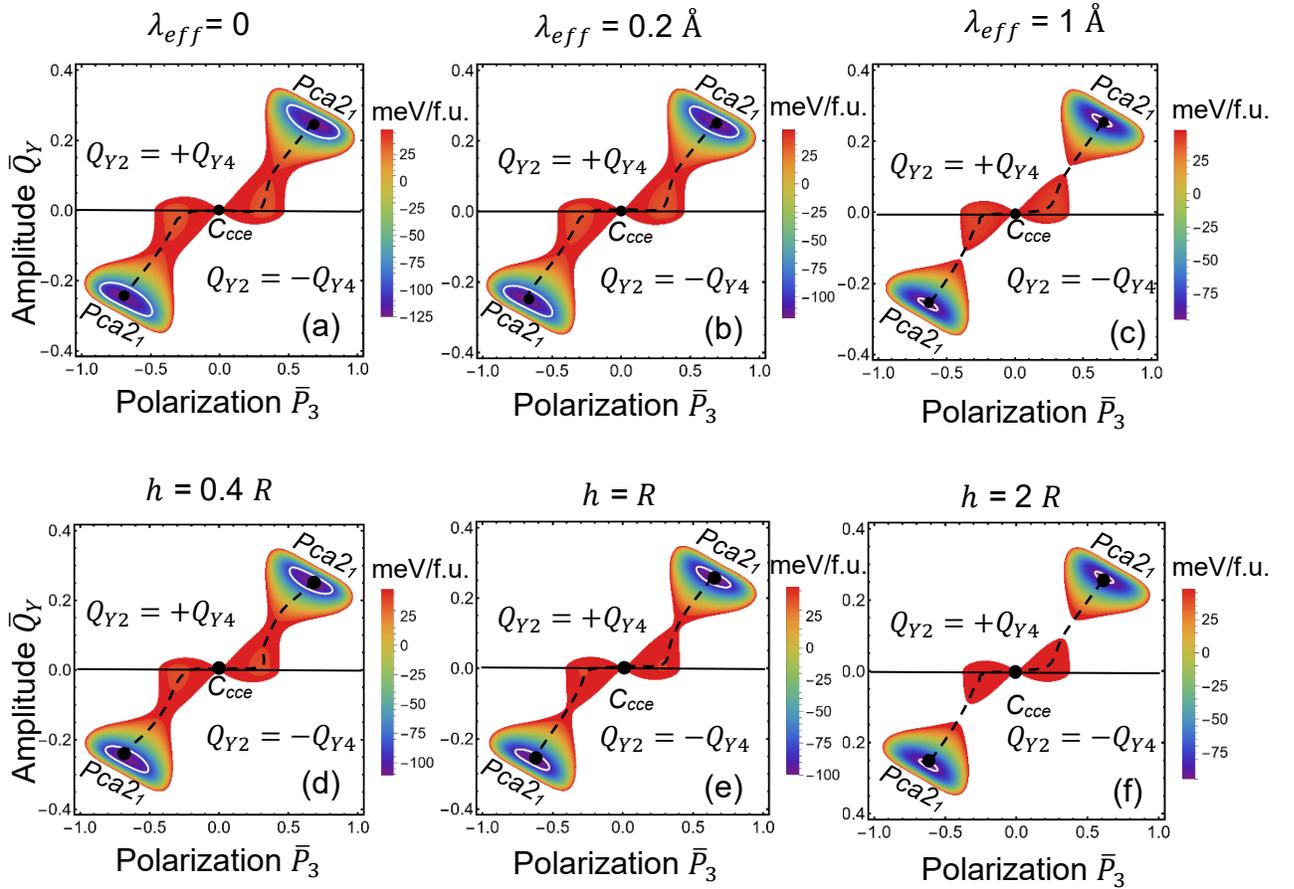

**FIGURE S3.** Free energy density as a function of $\bar{P}_3$ and $\bar{Q}_Y$, where either $\bar{Q}_Y = Q_{Y2} = Q_{Y4}$ (upper half-plane) or $\bar{Q}_Y = Q_{Y2} = -Q_{Y4}$ (bottom half-plane). The plots **(a)-(c)** are calculated for different screening lengths $\lambda_{eff} = 0$ **(a)**, 0.2 Å **(b)**, 1 Å **(c)**, $h =$10 nm and $u_m = -2.2$ %. The plots **(d)-(f)** are calculated for different radii $R = 25$ nm **(a)**, 10 nm **(b)**, 5 nm **(c)**, $h = 10$ nm, $\lambda_{eff} = 0.2$ Å and $u_m = -2.2$ %. White elliptic contours in the plots correspond to the energy of the bulk m-phase $f_m = -92$ meV/f.u. (counted from the t-phase). Dashed curves show the polarization switching path, which corresponds to the lowest energy barrier $f_{Ccce} = +48$ meV/f.u. (counted from the t-phase). Other parameters are $\Lambda_P = 10$ μm and $h_d = 10$ nm.